\documentclass[aps,showpacs]{revtex4} 

\usepackage{amssymb} 
\usepackage{graphicx} 
\usepackage[T1]{fontenc} 
\usepackage{euscript} 

\begin{document} 

\newcommand{\cequ}{\begin{eqnarray}} 
\newcommand{\fequ}{\end{eqnarray}} 
\newcommand{\anticomut}[2]{\left\{#1,#2\right\}} 
\newcommand{\comut}[2]{\left[#1,#2\right]} 
\newcommand{\comutd}[2]{\left[#1,#2\right]^{*}} 

\title{\bf 
Charged shells 
in Lovelock gravity: Hamiltonian treatment and physical implications} 
\author{Gon\c{c}alo A. S. Dias\footnote{Email: gadias@fisica.ist.utl.pt}} 
\affiliation{Centro Multidisciplinar de Astrof\'{\i}sica - CENTRA \\ 
Departamento de F\'{\i}sica, Instituto Superior T\'ecnico,\\ 
Universidade T\'ecnica de Lisboa,\\ 
Av. Rovisco Pais 1, 1049-001 Lisboa, Portugal} 
\author{Sijie Gao\footnote{Email: sijie@bnu.edu.cn}} 
\affiliation{Department of Physics, Beijing Normal University,\\ 
Beijing 100875, China} 
\author{Jos\'e P. S. Lemos\footnote{Email: lemos@fisica.ist.utl.pt}} 
\affiliation{Centro Multidisciplinar de {}Astrof\'{\i}sica - CENTRA \\ 
Departamento de F\'{\i}sica, Instituto Superior T\'ecnico,\\ 
Universidade T\'ecnica de Lisboa,\\ 
Av. Rovisco Pais 1, 1049-001 Lisboa, Portugal} 

\begin{abstract} 
Using a Hamiltonian treatment, charged thin shells, static 
and dynamic, in spherically 
symmetric spacetimes, containing black holes or other specific type 
of solutions, in $d$ dimensional Lovelock-Maxwell theory are 
studied. The free coefficients that appear in the Lovelock theory are 
chosen to obtain a sensible theory, with a negative cosmological 
constant appearing naturally. Using an ADM description, one then finds 
the Hamiltonian for the charged shell system.  Variation of the 
Hamiltonian with respect to the canonical coordinates and conjugate 
momenta, and the relevant Lagrange multipliers, yields the dynamic and 
constraint equations.  The vacuum solutions of these equations yield a 
division of the theory into two branches, namely $d-2k-1>0$ (which 
includes general relativity, Born-Infeld type theories, and other 
generic gravities) and $d-2k-1=0$ (which includes Chern-Simons type 
theories), where $k$ is the parameter giving the highest power of the 
curvature in the Lagrangian.  There appears an additional parameter 
$\chi=(-1)^{k+1}$, which gives the character of the vacuum solutions. 
For $\chi=1$ the solutions, being of the type found in general 
relativity, have a black hole character.  For $\chi=-1$ the solutions, 
being of a new type not found in general relativity, have a totally 
naked singularity character.  Since there is a negative cosmological 
constant, the spacetimes are asymptotically anti-de Sitter (AdS), and 
AdS when empty.  The integration from the interior to the exterior 
vacuum regions through the thin shell takes care of a smooth junction, 
showing the power of the method.  The subsequent analysis is divided 
into two cases: static charged thin shell configurations, and 
gravitationally collapsing charged dust shells (expanding shells are 
the time reversal of the collapsing shells).  In the collapsing case, 
into an initially non singular spacetime with generic character or an 
empty interior, it is proved that the cosmic censorship is definitely 
upheld.  Physical implications of the dynamics of such shells in a 
large extra dimension world scenario are also drawn.  One concludes 
that, if such a large extra dimension scenario is correct, one can 
extract enough information from the outcome of those collisions as to 
know, not only the actual dimension of spacetime, but also which 
particular Lovelock gravity, general relativity or any other, is the 
correct one at these scales, in brief, to know $d$ and $k$. 

\end{abstract} 


\maketitle 

\newpage 
\centerline{} 
\newpage 

\section{Introduction} 

In a world with extra large space dimensions, of the order of some 
microns or less, as postulated in \cite{arkani1}, the gravitational 
field sees all the dimensions, whereas the standard model fields are 
confined to the usual world, or brane, of three space dimensions. 
Thus, the four dimensional spacetime of the brane can be seen as 
embedded in the $d$ dimensional spacetime of the whole world.  In this 
setup several different problems are tackled. For instance, one 
solves, in a way, the hierarchy problem, since both the Planck scale 
and the electroweak scale can be made of the same order, and so 
quantum gravity, being a electroweak scale phenomenon, can be 
experimentally tested. Indeed, by smashing particles against each 
other in experiments with 
the new generation accelerators or in eventual cosmic ray 
collisions, one can produce black holes, or perhaps other spacetimes with 
different causal structure, with tiny masses and radii.  As this is 
a gravitational phenomenon, and the gravitational field spreads into 
all dimensions, the newly created spacetimes probe all the spacetime 
dimensions, and in addition render visible some quantum effects 
(see, e.g., \cite{arkani2,cavaglia,kanti,cardoso}). 

Now, since in this setting one might expose quantum gravity and large 
extra dimension phenomena, in order to study the higher dimensional 
black holes, or other spacetimes with different causal structure, 
formed in the collision of particles, one should consider studying 
these objects in possible natural extensions to the theory of general 
relativity, conceivably appropriate to a further quantum framework 
development. The action of general relativity in $d$ spacetime 
dimensions is proportional to the integral of the cosmological 
constant plus the Ricci scalar, and possible extensions should include 
powers of the Riemann tensor, such as the Kretschmann scalar, powers 
of the Ricci tensor, and powers of the Ricci scalar \cite{burgess}. In 
principle one has to choose a criterion to pick up the right 
combination of curvature scalars. For instance, adopting the criterion 
of keeping the same degrees of freedom, one may wonder what is the most 
natural generalization of general relativity for other dimensions. 
Such a generalization is given by the Lovelock action \cite{Lov}, 
where essentially one keeps the field equations second order.  In more 
detail, the Lovelock theory looks for an action which when properly 
varied yields symmetric tensors that are functions of the metric and 
its first and second spacetime derivatives, and divergence free.  In 
four dimensions the corresponding action is the Einstein-Hilbert 
action, proportional to the cosmological constant and the Ricci 
scalar. When varied, the Einstein-Hilbert action yields the Einstein 
tensor and the metric tensor, the only tensors that display the above 
properties and contribute to the equations of motion. In dimensions 
higher than four Lovelock found that only certain precise combinations 
of higher powers of the curvature scalars could enter in the action 
\cite{Lov}.  The interpretation in physical, geometric and topological 
terms of these precise combinations was put forward by Teitelboim and 
Zanelli \cite{teitelboimzanelli1} (see also 
\cite{teitelboimzanelli2,HTZ}). In this context, one can first argue 
that in zero spacetime dimensions, a spacetime point, there is a zero 
dimensional topological invariant which is a numerical constant, 
called the cosmological constant.  When this term is integrated in one 
or in two dimensions to form an action, i.e., when the term is 
dimensionally continued to the next odd or even dimension, it 
contributes to the equations of motion, which in this case are 
somewhat trivial, giving either a zero metric or a zero cosmological 
constant with an indeterminate metric.  Now, in two dimensions, there 
is also the corresponding topological invariant. This invariant, the 
Euler characteristic, is obtained by integrating the so called 
Gauss-Bonnet term. In two dimensions the Gauss-Bonnet term is the two 
dimensional Ricci scalar, which can thus be consistently considered 
the corresponding Euler density.  This term, being topological, does 
not contribute to the equations of motion in two dimensions.  However, 
when both these terms, the cosmological constant (the topological 
invariant in zero dimensions) and the Ricci scalar (the topological 
invariant in two dimensions) are integrated in three or in four 
dimensions, i.e., when they are dimensionally continued to the next 
odd or even dimensions, they contribute non-trivially to the equations 
of motion, and indeed in three and four dimensions, yield the standard 
general relativity. Now, given general relativity in four dimensions, 
one can add a generalized Gauss-Bonnet (i.e., a generalized Euler 
density) term, in four dimensions, first discovered by Lanczos 
\cite{Lanc}, which when integrated in the four dimensional action 
gives essentially the corresponding Euler characteristic, also a 
topological invariant not contributing to the classical field 
equations. In turn, in five and six dimensions, not only an action 
with the cosmological constant and the Ricci scalar, mentioned above, 
contribute to the equations of motion, but also one can dimensionally 
continue the previous generalized Gauss-Bonnet term in order to have a 
meaningful extended action, which gives the desired equations of 
motion of second order.  Of course, in six dimensions there is a new 
generalized Gauss-Bonnet term, which when integrated gives the 
corresponding Euler characteristic, not contributing to the equations 
of motion.  Then, repeating this process of adding a generalized 
Gauss-Bonnet term of the previous even dimension to the next two 
dimensions, odd and then even, one gets the Lovelock gravity of that 
specific dimension.  Thus, the Lovelock theory can also be considered 
as a dimensional continuation of the Euler characteristics of lower 
dimensions \cite{teitelboimzanelli1,teitelboimzanelli2,HTZ}.  The 
theory has, in addition to Newton's constant and the cosmological 
constant, new arbitrary dimensionful parameters, which 
should be chosen carefully.  We have given a classical motivation for 
including certain special new terms in the gravitational action in $d$ 
dimensions, but there are also quantum motivations to study the 
dimensional continuation of the Euler characteristics.  Indeed, in a 
string theory context, in order to prevent the existence of ghosts in 
a low energy limit of the theory, one has to add to the $d$ 
dimensional general relativity action all the previous lower 
dimensional Euler characteristics \cite{Zwie,Zumi}. 

Suppose now the world has indeed extra dimensions, and that the whole 
bulk spacetime obeys Lovelock gravity rather than Einstein 
gravity. Thus, in this case, it is important to study generic 
properties of black holes in the Lovelock theory. Solutions of 
spherical black holes in $d$ spacetime dimensions for Einstein-Hilbert 
action plus a generalized Gauss-Bonnet term, i.e., a specially 
truncated Lovelock theory, were found in 
\cite{bolwaredeser,Whee,Whitt,MSW,wiltshire}. Thermodynamics were 
often studied in these works.  In \cite{BTZ1}, where black hole 
solutions were also found, no relevant Lovelock term was put to zero, 
rather a special choice was made for all the Lovelock coefficients, 
which depend only on the cosmological constant and the dimension of 
the spacetime. All non-topological terms are included in the action, 
and the theory can thus be called dimensionally continued gravity. 
This choice is one among infinite, but it leads to a natural outcome, 
where the action in even dimensions has a Born-Infeld form, i.e., it 
may be regarded as the gravitational analogue of the Born-Infeld 
electrodynamics, and in odd dimensions has a Chern-Simons form.  Also 
in \cite{BTZ1}, Lovelock theory was coupled to Maxwell 
electromagnetism, and the corresponding black hole solutions with 
electric charge and cosmological constant were found.  Now, a natural 
extension of this prescription was advanced in \cite{CTZ}, where it 
was allowed that the topological term plus some non-topological terms, 
starting from the top, could be put to zero. For instance, in $d=10$ 
dimensions in this setting one may work just with the general 
relativity term and the following generalized Gauss-Bonnet term, by 
putting to zero the other higher dimensional Euler densities.  For the 
nonzero terms, the choice of the coefficients of the theory is the 
same as in \cite{BTZ1}. This extension has as a special case the 
dimensionally continued gravity of \cite{BTZ1}.  By coupling this 
extension to Maxwell electromagnetism, a general class of charged 
vacuum spacetimes with a negative cosmological constant, including of 
course the black holes found in \cite{BTZ1}, were exhibited 
\cite{CTZ}. 
These vacuum solutions 
yield a natural division of the Lovelock theory into two branches, 
namely $d-2k-1>0$ (which includes general relativity, Born-Infeld 
type theories, and other generic gravities) 
and $d-2k-1=0$ (which includes Chern-Simons type theories), 
where $k$ is the parameter that 
gives the highest power of the curvature in the Lagrangian. 
In the solutions there 
appears an additional parameter $\chi$, with $\chi=(-1)^{k+1}$. 
This implies that, in addition, the solutions can be subdivided 
into two families of different characters, one family $\chi=1$, comprising 
solutions of the type found in general relativity, has a black hole 
character (meaning that for a correct choice of the parameters, such 
as mass and charge, the solution is a black hole solution, although, 
of course, for other choices of parameters it can be an extremal black 
hole or a naked singularity), and the other family $\chi=-1$, 
being of a new type not 
found in general relativity, has a naked singularity character 
(meaning there is no possible choice of the parameters that gives a 
black hole solution, the full vacuum solution is always singular 
without horizons).  Of course the solutions also include empty 
spacetimes. Since in general there is a negative cosmological constant 
the spacetimes are asymptotically anti de Sitter (AdS), or when empty 
they are AdS itself or some form of it. 

Given that one has a family of spacetimes with either black hole or 
naked singularity character, it is now important to study matter 
effects, either static or dynamic (collapsing or expanding), on these 
solutions. Such a phenomenon could be interesting in the aftermath of 
a collision between charged particles where the debris formed in the 
collision can be accreted or excreted in the newly formed charged 
spacetime.  Accretion and excretion processes are usually technically 
elaborate, so to turn the analysis simpler one can study the case of a 
thin shell in the background spacetime, be it a black hole or 
otherwise.  Static and dynamic solutions on such backgrounds are then 
of interest. 
However, in Lovelock theories, even a thin shell and its 
dynamics can bring complications. Indeed, in general relativity for 
instance, to study a thin shell in a given background one has to make 
a smooth junction from the interior to the exterior spacetime, as was 
done in \cite{Isr}. Now, the junction conditions, in general, depend 
on the theory one is studying, and for theories with higher powers on 
the Riemann and other tensors this can be nontrivial \cite{davis}.  An 
elegant and useful approach, that bypasses in a way the junction 
conditions for a spacetime with a thin shell, uses a Hamiltonian 
formalism for the theory under study. 
This approach was developed by 
H\'{a}j\'{\i}\v{c}ek and Kijowski \cite{hajicekkijowski} for the 
theory of general relativity with matter 
in four dimensions, and was subsequently 
explored by Cris\'ostomo and Olea in $d$ dimensional general 
relativity coupled to Maxwell theory and matter 
with applications mainly in three 
spacetime dimensions \cite{CriOl,Ol}.  The method requires that one 
puts the theory in a Hamiltonian form and then one directly integrates 
the canonical constraints, producing the shell dynamics for the 
background one wants to study.  The pay off is that, not only one 
avoids to have to develop covariant junction conditions for each 
particular theory one is studying, but moreover the full spacetime, 
comprised of interior, shell, and exterior components, is treated as 
whole. This method is particularly well suited for symmetric 
configurations. The great power of the formalism was brought up in 
\cite{CCS} where it was applied in pure Lovelock gravity to thin shell 
collapse in the background of the uncharged spacetimes 
found in \cite{CTZ}. 
Now, since the newly formed spacetime, black hole or otherwise, 
generated from a collision between charged particles, is most probably 
charged, and for the same reason the collapsing or expanding debris 
are also charged, it is of interest to study, using the 
Hamiltonian formalism advanced in \cite{CCS}, the gravitational 
dynamics of a charged thin shell in the charged sector of the 
spacetime solutions, black hole or otherwise, found in \cite{CTZ}, of 
Lovelock gravity coupled to Maxwell electromagnetism. 
In the usual extra dimension scenarios the electromagnetic and matter 
fields, being confined to the brane, do not probe the extra 
dimensions, so that an axisymmetric shell, static or collapsing, 
around a spherically symmetric black hole is an important 
configuration to study. However, in order to simplify the analysis we 
study instead, as a zero order approximation, a spherically symmetric 
shell around a spherically symmetric black hole (such a configuration 
would be quite realistic for charged fields and particles that can 
probe the extra dimensions). 
Of course, within general relativity, a 
very special case of Lovelock gravity, such an analysis should recover 
the earlier results, obtained through a junction condition formalism 
\cite{Isr} of charged shell dynamics in a charged black hole 
spacetime \cite{Kuch,Boul,GL}. 
All the works just mentioned are classical, and an extension of 
full Lovelock gravity into the 
quantum domain, or even within a semiclassical approximation, although 
important, seems beyond reach. 

In the present article we extend the treatment of the classical 
dynamics of an uncharged thin shell in an uncharged spacetime, black 
hole or otherwise, with a negative cosmological constant background in 
Lovelock-Maxwell theory, given in \cite{CCS}, to the classical 
dynamics of a charged thin shell in a charged spacetime, black hole or 
otherwise, with a negative cosmological constant background in 
Lovelock-Maxwell theory, and among other things, show the power of the 
Hamiltonian method in unifying in a single unit the three sectors of 
the problem, namely, the gravitational, the electrodynamic and the 
matter sectors, as well as taking into account 
automatically the smooth junction of the several regions of spacetime in 
question, 
the interior, 
the exterior and the thin shell in between. 
In Sec. \ref{LLSection} the system under study, namely charged vacuum 
plus charged thin shell, is initially put in an action and Lagrangian 
framework, in both tensor and differential forms languages.  The field 
content of the action is composed of three sectors, the gravitational, 
the electrodynamic, and the matter term.  In the gravitational action 
we establish the coupling constants and make the choice of Lovelock 
coefficients, cutting off the Lovelock polynomial at the highest 
possible power of the curvature for which the theory is sensible.  In 
the electrodynamic action we write the Maxwell term and add a current 
term, which describes the current of charged matter in the shell, and 
precede each term with its coupling constant.  In the matter action we 
define the energy-momentum tensor of a perfect fluid matter source. 
We then write the same action in the Hamiltonian form in tensorial 
language.  For each of the above mentioned sectors of the action we 
define the canonical coordinates and its conjugate momenta, up to 
surface terms.  As the system under study is a system with 
constraints, we write the constraints and their respective Lagrange 
multipliers for each of the sectors of the action.  Next, we obtain 
the Hamiltonian field equations from the Hamiltonian action by using 
spherically symmetry in the action and varying it with respect to the 
canonical coordinates and momenta, for the evolution equations, and 
the Lagrange multipliers, for the constraint equations. 
In Sec. \ref{solutions} we analyze the vacuum solutions of the derived 
Hamiltonian equations and describe their properties.  Afterwards we 
establish the geometrical framework of the thin shell and its 
Hamiltonian description. We divide spacetime into three parts, 
interior spacetime, thin shell, and exterior spacetime, and describe 
their geometric set-up. Then we specify the matter properties of the 
thin shell, which we define as being that of a perfect fluid. We write 
its energy-momentum tensor and its relevant projections.  We then 
solve the complete Hamiltonian equations around the thin shell. 
First, we note that the vacuum solutions 
yield a natural division of the Lovelock theory into two branches, 
namely $d-2k-1>0$ and $d-2k-1=0$, where $k$ is the parameter that 
gives the highest power of the curvature in the Lagrangian. There 
appears an additional parameter $\chi$, with $\chi=(-1)^{k+1}$, which 
gives the character of the solutions, namely, the vacuum solutions may 
have a black hole character $(\chi=1)$ or a naked singularity character 
($\chi=-1$).  Since there is a 
negative cosmological constant, the spacetimes are asymptotically 
AdS, and AdS  when 
empty, or some form of it. 
Second, the whole integration of the equations from the interior to 
the exterior vacuum regions through the thin shell is performed, 
where the smooth junction of the several regions of spacetime in 
question is automatically taken into account by the integration, 
showing definitely the efficacy of the method. 
The 
shell's dynamics in the vacuum spacetimes 
is then derived and the electrodynamic constraint 
equation recovers the electric charge conservation. 
An analysis of the equations of the thin shell is performed by 
studying two interesting cases, namely a static thin shell in 
equilibrium and the gravitational collapse of a thin shell (the study 
of gravitational expansion is simply the time reversal of 
gravitational collapse, and so it is not necessary to analyze it in 
any detail). For the static shell we determine the radii at which the 
shell is in equilibrium and the pressure necessary to maintain the 
shell at this radius. For the gravitational collapse of a thin shell 
we start studying a simple example of dust matter in an empty interior 
and then prove cosmic censorship in a general case.  More 
specifically, following the natural division of the 
Lovelock theory into the two branches mentioned above, 
we study the collapse of a thin shell onto an empty 
interior without cosmological constant, and give as examples collapse 
in four and ten dimensions in general relativity, collapse in ten 
dimensions in Born-Infeld type theories, and collapse in ten 
dimensions in general relativity with a single generalized 
Gauss-Bonnet term. We also study gravitational collapse in five 
dimensions onto an empty interior with a cosmological constant as an 
example of Chern-Simons type theories.  Plots are drawn for these 
cases just mentioned, which illustrate the dynamics of the collapse 
and exhibit the cosmic censorship at work.  This example of collapse 
of a shell into an empty interior, in a number of theories specified 
by $d$ and $k$, shows that electric charge provides a mechanism for 
cosmic censorship for all relevant cases, including those which, in 
the uncharged case, did not conform to it.  Then, more generally, for 
gravitational collapse onto black hole interiors, it is proven that 
the equations respect the cosmic censorship hypothesis. 
In Sec. \ref{section4} we conclude and draw some 
physical implications in connection 
with the extra large dimensions scenario.  We put $c=1$. 

\section{Action, Lagrangian, Hamiltonian and equations of 
motion in Lovelock gravity coupled to Maxwell electrodynamics 
and a charged thin shell in a spherically symmetric 
background} 
\label{LLSection} 

\subsection{Action, Lagrangian, and Hamiltonian in a general form} 

\subsubsection{Action and Lagrangian} 

For our purposes of studying charged matter in a $d$ dimensional charged 
background spacetime, the field content is divided into three sectors, 
the gravitational, the electrodynamic, and the matter sectors.  Then, 
the action $I$ can be written as the sum of a gravitational action 
$I^{(g)}$, an electrodynamic action $I^{(e)}$, and a generic matter 
field action $I^{(m)}$, which will be specialized later on to be a 
charged thin shell. Thus, 
\begin{equation} 
I=I^{(g)}+I^{(e)}+I^{(m)}\,. 
\label{actiontotal} 
\end{equation} 

\vskip 0.2cm 
\noindent {The gravitational action and Lagrangian:} 
\vskip 0.1cm 
\noindent 
The gravitational sector of the action, $I^{(g)}$, depends on 
which theory one wants to adopt.  General relativity, first formulated 
in four spacetime dimensions, can be trivially extended to higher $d$ 
spacetime 
dimensions by changing the action from a four dimensional integral of 
the cosmological constant term and the Ricci scalar, to a $d$ 
dimensional integral of both terms.  However, this extension is no 
longer unique. Another natural extension is given by the Lovelock 
gravity \cite{Lov}, in which its action results from demanding that 
the Euler-Lagrange equations derived from the corresponding 
Lagrangian yield all tensors $A^{\mu\nu}$ symmetric in 
$\mu\nu$, occurring concurrently with the metric tensor 
$g_{\mu\nu}$ and its first two derivatives, i. e., 
$A^{\mu\nu}=A^{\mu\nu}(g_{\rho\sigma};g_{\rho\sigma;\gamma}; 
g_{\rho\sigma;\gamma\delta})$, and divergence free.  In four dimensions 
only the cosmological constant and the Ricci scalar yield an action 
with these properties, which is precisely the Einstein-Hilbert 
action. As one goes to higher dimensions new generalized Gauss-Bonnet 
terms, topological in nature in the previous dimension, make their 
appearance.  As mentioned in the Introduction, the Lovelock theory can 
then be considered as a dimensional continuation of the topological 
Euler characteristics of lower dimensions 
\cite{teitelboimzanelli1,teitelboimzanelli2,HTZ}.  The theory has, in 
addition to Newton's constant and the cosmological constant, new 
arbitrary dimensionful parameters, which should be chosen carefully. 
Due to its interest and generality, we work here with Lovelock 
gravity. The gravitational part $I^{(g)}$ in (\ref{actiontotal}) is then 
the Lovelock action, which turns out to be a polynomial in the 
curvature tensor, of degree $[d/2]$, where the brackets $[d/2]$ 
represent the integer part of $d/2$. 
This action, and the corresponding Lagrangian $\cal L$ 
defined through $I=\int d^d\,x\,{\cal L}$, can be written as 
\begin{equation} 
I^{(g)}=\kappa \int_{{}_{\cal M}}d^dx 
\sum_{p\,=\,0}^{[d/2]}\alpha_p\,\sqrt{-\mathfrak{g}}\,2^{-p}\, 
\delta^{\mu_1\cdots \mu_{2p}}_{\nu_1\cdots 
\nu_{2p}}\,R_{\mu_1 \mu_2}^{\nu_1\nu_2} \cdots 
R_{\mu_{2p-1} \mu_{2p}}^{\nu_{2p-1}\nu_{2p}}\,, 
\label{llactioncomponent form} 
\end{equation} 
where $\kappa$ is inversely proportional to the 
Newton's constant (which will be appropriately chosen below), 
$\cal M$ 
stands for the spacetime manifold, 
$\mathfrak{g}$ is the determinant of the spacetime metric, and 
$\mu_1,\mu_2, ...$ are spacetime indices. 
The generalized $\delta$ 
function is antisymmetric in all of the upper indices and all of the 
lower indices, and $R_{\mu_1 \mu_2}^{\nu_1\nu_2}$ 
is the Riemann tensor. 
The coefficients $\alpha_p$ are arbitrary in general, apart from 
the first two $\alpha_0$ and $\alpha_1$. 
Indeed, in the gravitational action (\ref{llactioncomponent form}), 
the first term of the integrand 
is the cosmological constant $\Lambda$ and the 
second term of the integrand is the 
Ricci scalar $R$, i.e., the terms of the Einstein-Hilbert action. This 
shows that general relativity is contained in the Lovelock theory as a 
particular case, namely by putting in the action all the $\alpha_p=0$ 
for $p\geq2$.  For even dimensions, the term $p=d/2$ in the action 
(\ref{llactioncomponent form}) is the Euler 
characteristic of that $d$ dimensional manifold and does not 
contribute to the field equations. However, although not relevant for 
the purposes of this paper, the presence of the Euler term guarantees 
the existence of a well defined variational principle for 
asymptotically locally AdS spacetimes (see \cite{CTZ}).  Lovelock 
gravity, from the way it is constructed, has the same essence as 
general relativity. However, the theory is more complicated, it sets 
extra problems not present in general relativity and yields new 
interesting features.  For instance, for $d>4$ the fields may evolve 
in a non-unique manner, such that, given initial values for the fields 
at $t=t_0$, at $t>t_0$ the equations of motion do not determine those 
fields completely. This is due to the presence in the Lagrangian of 
high powers in the first derivative of the metric tensor with respect 
to time \cite{teitelboimzanelli1,teitelboimzanelli2,HTZ}.  An 
interesting aspect of Lovelock gravity is that, although its 
linearized approximation is classically equal to the corresponding 
linearized approximation in general relativity \cite{Zumi}, in the 
full strong gravity regime of the theory, the higher powers of 
curvature in the Lagrangian yield solutions that are different, and 
such type of solutions 
in Lovelock theory cannot be reached through a solution in 
general relativity \cite{Whee}-\cite{CCS}. 

                   Another feature is the fact that, in addition to the 
constant 
$\kappa$ (inversely proportional to an appropriate 
generalization of Newton's 
constant) and the cosmological constant $\Lambda$, there are the 
$[(d+1)/2]$ arbitrary dimensionful parameters. Now, for a given 
dimension and an arbitrary choice of the coefficients $\alpha_p$, the 
dynamical evolution can become unpredictable 
\cite{teitelboimzanelli1,teitelboimzanelli2,HTZ}, thus it is 
advantageous to restrict these coefficients in order to be able to 
construct meaningful black hole and other solutions. An important step 
in this direction was taken first in \cite{BTZ1} and generalized in 
\cite{CTZ}, where the coefficients $\alpha_p$ were chosen so that 
unique sensible solutions, with well defined perturbations could be 
found. One way to meet these requirements is to demand, in 
a consistent way, that the theory has 
a unique cosmological constant $\Lambda$ 
as shown in \cite{BTZ1,CTZ}, which 
leads to the choice 
\begin{eqnarray} 
\alpha_p\,&:=&\,c_p^k\,=\, \left\{ 
\begin{array}{ll}\frac{l^{2(p-k)}}{(d-2p)} \left(\begin{array}{l} k \\ 
p \end{array}\right), & p\leq k \\0, & p>k \end{array}\right.\,, 
\label{alphaconstants} 
\end{eqnarray} 
where $l$ is a length scale of the theory given in 
terms of the cosmological constant $\Lambda$ by 
\begin{equation} \Lambda=-\frac{(d-1)(d-2)}{2l^2}\,. 
\end{equation} 
As seen in the last expression, in this setting the 
cosmological constant is negative throughout, and so the solutions 
will be asymptotically AdS solutions.  We are then left 
with a set of theories labeled by an integer $k$, with 
$1\,\leq\,k\,\leq\,[(d-1)/2]$.  One should explain better the 
dependence of the theories on $k$.  In \cite{BTZ1} $k$ was put equal 
to its maximum value always, $k=[(d-1)/2]$, i.e., no pertinent 
Lovelock term is put to zero, and one has the corresponding special 
choice made in Eq. (\ref{alphaconstants}).  In this case, all 
non-topological terms, plus the topological non-dynamical term in the 
even dimensional case, are included in the action, and the theory can 
thus be called dimensionally continued gravity, since as one goes one 
dimension up, what was a topological term in the previous dimension, is 
now a term that contributes dynamically when one continues the 
dimension.  In even dimensions, $d\geq4$, this choice leads to a 
Born-Infeld gravitational action, the gravitational analogue of the 
Born-Infeld electrodynamics, and in odd dimensions, $d\geq5$, it leads 
to a Chern-Simons gravitational action \cite{BTZ1}. 
This dimensionally continued 
theory \cite{BTZ1} can be naturally extended by canceling higher 
order terms, from a given integer $k$ upwards. This was 
done in \cite{CTZ}. 
For instance one can 
set up a theory in which only the terms $p=0$ and $p=1$ appear, thus 
$k=1$. This theory is general relativity. If one puts $k=2$ one gets 
general relativity plus the corresponding generalized Gauss-Bonnet 
term, and so on, up to the dimensionally continued gravity where 
$k=[(d-1)/2]$. In brief, for a given dimension $d$, depending on the 
choice of the integer $k$, one generates different theories, with $k$ 
giving the highest power of the curvature in the Lagrangian.  Now, the 
other fundamental constant $\kappa$ can be related to a generalized 
Newton's constant $G_k$, labeled by $k$, with units $[G_k]=({\rm 
length})^{d-2k}$, through 
\begin{equation} 
\kappa=\frac{1}{2(d-2)!\Omega_{(d-2)}G_k}\,, 
\label{kappa} 
\end{equation} 
where $\Omega_{d-2}$ is the area of the unit 
sphere 
in $d-2$ dimensions. These theories have well defined black hole 
configurations. 

\vskip 0.2cm 
\noindent {The electrodynamic action and Lagrangian:} \vskip 0.1cm 
\noindent 
The electrodynamic sector of the action 
(\ref{actiontotal}), the electrodynamic action $I^{(e)}$, can be chosen 
to be the electromagnetic 
Maxwell term plus a matter electric current, and can be 
written as 
\begin{equation} I^{(e)}=-\frac{1}{4 \epsilon \Omega_{d-2}}\int_{\cal 
M} d^d\,x\;\sqrt{-\mathfrak{g}}\, F_{\mu\nu}F^{\mu\nu} 
-\frac{1}{\epsilon}\int_{\cal M} d^d\,x\; \sqrt{-\mathfrak{g}}\, J^\mu 
A_\mu\,, 
\label{maxwellterm2} 
\end{equation} 
where $\epsilon$ is related to the vacuum permitivity 
$\epsilon_0$ through the expression 
$\epsilon=\frac{\epsilon_0}{\Omega_{d-2}}$, 
$\mathfrak{g}$ is the determinant of the spacetime metric 
$\mathfrak{g}_{\mu\nu}$, $F^2=F_{\mu\nu}F^{\mu\nu}$ is the square of 
$F_{\mu\nu}=\nabla_\mu A_\nu-\nabla_\nu A_\mu$, which 
is the Maxwell tensor, with $A_\mu$ being the electromagnetic 
potential, and $J^\mu$ is the electric current.  For vacuum solutions 
$J^\mu$ is zero, and one has pure 
Maxwell electromagnetism. 

\vskip 0.2cm 
\noindent {The matter action and Lagrangian:} 
\vskip 0.1cm 
\noindent 
The last term in the action (\ref{actiontotal}), the matter action 
$I^{(m)}$ is defined as 
\begin{equation} I^{(m)}=\int\, d^dx \,\sqrt{-\mathfrak{g}} \,{\cal 
L}_m\,, 
\label{matterterm} 
\end{equation} 
where the matter Lagrangian ${\cal L}_m$ is defined 
through the energy-momentum tensor of the matter $T_{\mu\nu}$ by 
\begin{equation} 
T_{\mu\nu}=-\frac{(d-2)!\,\Omega_{(d-2)}}{4\,\pi\,\sqrt{-\mathfrak{g}}} 
\frac{\delta I^{(m)}}{\delta g^{\mu\nu}}\,, 
\label{energymomentumtensor} 
\end{equation} 
where $\delta$ here means variation. In vacuum 
$I^{(m)}=0$. Later on we will specify the matter as being 
made of a charged 
thin shell. 

\subsubsection{Hamiltonian} 

In order to apply the 
Hamiltonian framework, we use the Arnowitt-Deser-Misner (ADM) 
formulation \cite{ADM} where there is a foliation of spacetime into 
$t=$ constant hypersurfaces, denoted by $\Sigma_t$. In this foliation 
of the spacetime the metric, both inside and outside, is written 
generically as 
\begin{equation} 
ds^2=-(N^{\bot})^2dt^2+g_{ij}(N^idt+dx^i)(N^jdt+dx^j)\,,\label{ADMmetric} 
\end{equation} 
where $N^{\bot}$ and $N^i$ are the lapse and shift 
functions of the foliation, respectively. The $g_{ij}$ are the metric 
coefficients of the intrinsic geometry of the hypersurfaces 
$\Sigma_{t}$, where $i,j$ run only on the spatial components. 
It is known that the action $I=\int d^dx\,{\cal L}$, 
with ${\cal L}$ the Lagrangian, can be 
written as 
\begin{equation} 
I=\int dt\, \int d^{d-1}x\left( 
\pi^{ij}\dot{g_{ij}}-{\cal H}\right)\,, 
\label{actionhamiltonian} 
\end{equation} 
where $\pi^{ij}=\delta {\cal L}/\delta \dot{g_{ij}}$ 
are the momentum components conjugate to the metric 
components $g_{ij}$, 
$\dot{g_{ij}}$ are the respective coordinate time derivatives, and 
${\cal H}$ is the Hamiltonian of the system. 
It is 
then useful to write the Hamiltonian in the form 
\begin{equation} 
\mathcal{H}=N^{\bot}\mathcal{H}_\bot+N^i\mathcal{H}_i\,, 
\label{hamiltonian} 
\end{equation} 
where $\mathcal{H}_\bot$ is the normal Hamiltonian 
constraint that generates the time 
translations normal to the hypersurface $\Sigma_t$, and 
$\mathcal{H}_i$ are the tangential constraints 
that generate the translations in each of the 
hypersurfaces $\Sigma_t$, which is the same as saying that 
$\mathcal{H}_i$ is the generator of hypersurface diffeomorphisms, or 
that it generates coordinate transformations in $\Sigma_t$. 
In addition, one can write, 
\begin{equation} \mathcal{H}_{\bot}=\mathcal{H}_{\bot}^{(g)}+ 
\mathcal{H}_{\bot}^{(e)}+\mathcal{H}_{\bot}^{(m)}\,,\quad 
\mathcal{H}_{i}=\mathcal{H}_{i}^{(g)}+ \mathcal{H}_{i}^{(e)}+ 
\mathcal{H}_{i}^{(m)}\, 
\label{hamiltoniancomponents} 
\end{equation} 
where the superscript $(g)$ refers to the 
gravitational sector, the superscript $(e)$ refers to the 
electrodynamic sector, and the superscript $(m)$ refers to the 
matter sector of the Hamiltonian. The lapse function $N^{\bot}$ and the 
shift 
functions $N^i$ of the ADM metric act here as Lagrange multipliers. 

\vskip 0.2cm 
\noindent {The gravitational Hamiltonian:} 
\vskip 0.1cm 

\noindent For the Lovelock theory, the Hamiltonian components of the 
gravitational field can be taken from the action 
(\ref{llactioncomponent form}), the momenta conjugate 
to the corresponding Lagrangian 
$\pi^{ij}=\delta {\cal L}/\delta \dot{g_{ij}}$, 
and 
$\dot{g_{ij}}$, with the result 
\cite{teitelboimzanelli1,teitelboimzanelli2,HTZ} 
\begin{eqnarray} \mathcal{H}_{\bot}^{(g)}&=&-\kappa \sqrt{g} \sum_p 
\frac{(d-2p)!\,\alpha_p}{2p}\, \delta^{i_1\cdots i_{2p}}_{j_1\cdots 
j_{2p}}\,R_{i_1 i_2}^{j_1j_2} \cdots R_{i_{2p-1} 
i_{2p}}^{j_{2p-1}j_{2p}}\,, 
\label{normalgravitationalhamiltonian}\\ 
\mathcal{H}_i^{(g)}&=&-2\pi^j_{i|j}\,. 
\label{surfacegravitationalhamiltonian} 
\end{eqnarray} 
The $R_{ijkl}$ in 
(\ref{normalgravitationalhamiltonian}) are the spatial components of 
the curvature tensor in the $d-$dimensional spacetime. The 
Gauss-Codazzi equations give us the relation between this tensor and 
$\hat{R}_{ijkl}$, the Riemann tensor intrinsic to the surface 
$\Sigma_t$, 
\begin{equation} 
R_{ijkl}=\hat{R}_{ijkl}+K_{ik}K_{jl}-K_{il}K_{jk}\,, 
\label{gausscodazzi} 
\end{equation} where $K_{ij}$ is the extrinsic curvature tensor of 
the surface, given by the expression, 
\begin{equation} K_{ij}=\left( \frac{1}{2N^\bot} \right) 
\left(N_{i;j}+N_{j;i}-\dot{g}_{ij} 
\right)\,. \label{extrinsiccurvature} 
\end{equation} The semicolon in the indices denotes the intrinsic 
covariant derivative in the hypersurface $\Sigma_t$, and the dot is 
the derivative of the spatial-spatial components of the metric with 
respect to the time coordinate.  The conjugate momentum to $g_{ij}$ 
can be found through $\delta {\cal L}/\delta \dot{g_{ij}}$, which in 
this case gives \cite{teitelboimzanelli1,teitelboimzanelli2,HTZ} 
\begin{eqnarray} \pi^i_j&=&-\kappa\sqrt{g}\sum_p 
\frac{p!\,(d-2p)!\,\alpha_p}{2^{p+1}}\sum^{p-1}_{s=0} D_{s(p)}\, 
\delta^{ii_1\cdots i_{2s} \cdots i_{2p-1}}_{jj_1\cdots j_{2s}\cdots 
j_{2p-1}}\times \nonumber \\ &\times& R_{i_1i_2}^{j_1j_2}\cdots 
R_{i_{2s-1}i_{2s}}^{j_{2s-1}j_{2s}}\,K_{i_{2s+1}}^{j_{2s+1}} \cdots 
K_{i_{2p-1}}^{j_{2p-1}}\,, 
\end{eqnarray} where, 
\begin{equation} D_{s(p)}= \frac{(-4)^{p-2}}{s!\,[s(p-s)-1]!!}\,. 
\label{dfunction} 
\end{equation} 
Here the double factorial is defined by: 
\begin{eqnarray} s!!&\equiv& \left\{ \begin{array}{cc} s\cdot (s-2) 
\ldots 5\cdot3\cdot1 & s>0 \,\, \textrm{odd}\\ s\cdot (s-2)\ldots 
6\cdot4\cdot2 & s>0 \,\, \textrm{even}\\ 1 & s=-1,0 
\end{array}\right.\,. 
\label{doublefactorial} 
\end{eqnarray} 

\vskip 0.2cm 
\noindent {The electrodynamic Hamiltonian:} \vskip 0.1cm 
\noindent 
The action in a Hamiltonian form of the electrodynamic field in a 
curved background is given by 
$ 
I_{e} 
= \int dt \, \int d^{D-1}x\, \left[ 
p^i\dot{A}_i-\frac12 N^{\bot} \left( \Omega_{d-2} \frac{1}{\sqrt{g}} p^i 
p_i 
+ \frac{\sqrt{g}}{2\Omega_{d-2}} F^{ij}F_{ij} \right) +\varphi p^i_{,i} 
-\varphi\,J^0\right]\nonumber \,, 
\label{emhamiltonianaction} 
$ 
where $p^i$ is the momentum conjugate to the spatial 
components of the gauge field $A_i$, $\varphi\equiv A_0$, $F_{ij}$ are 
the spatial components of the Maxwell tensor, $J^0$ is the time 
component of the electric current, and $\Omega_{d-2}$ is 
the area of the $(d-2)$-dimensional unit sphere. 
We are ignoring surface terms since they will be automatically 
taken into account in this setting. From the electrodynamic action, and 
using the definition given in Eqs. (\ref{hamiltonian}) and 
(\ref{hamiltoniancomponents}), we have 
\begin{eqnarray} 
\mathcal{H}^{(e)}_\bot &=& \frac12 \Omega_{d-2} 
\frac{1}{\sqrt{g}} p^i p_i + \frac{\sqrt{g}}{2\Omega_{d-2}} 
F^{ij}F_{ij}\,, 
\label{emnormalconstraint}\\ 
\mathcal{H}^{(e)}_i &=& 0\,. 
\label{emtangentialconstraint} 
\end{eqnarray} 
Besides the usual constraints, there is also a new 
constraint $E_\varphi$, associated with the Lagrangian multiplier 
$\varphi$ 
\begin{equation} E_\varphi\equiv p^i_{,i}-J^0 \,. 
\label{emconstraint} 
\end{equation} 

\vskip 0.2cm 
\noindent {The matter Hamiltonian:} 
\vskip 0.1cm 
\noindent 
The matter sector of the Hamiltonian constraint is 
written as 
\begin{eqnarray} 
\mathcal{H}_\bot^{(m)}&=&\sqrt{g}\,T_{\bot\bot}\,,\\ 
\mathcal{H}_i^{(m)}&=& 2\,\sqrt{g}\,T_{\bot i}\,, 
\label{matterconstraint} 
\end{eqnarray} 
where $T_{\mu\nu}$ is the matter energy-momentum tensor, and where the 
index $\bot$ means that the tensor has been contracted with the 
hypersurface normal $n_{\mu}=(-N^\bot,0,0,0)$. 
In the following we will specify the matter as being 
made of a charged 
thin shell. 

\subsection{Hamiltonian and field equations for  thin shells 
with interior and exterior static vacua in spherically symmetric 
spacetimes} 

\subsubsection{Gravitational, electrodynamic, and thin shell 
Hamiltonians in spherically symmetric spacetimes} Using the 
Hamiltonian formalism, we now want to set up the field equations 
appropriate for charged thin shells in static spherically symmetric 
charged Lovelock backgrounds.  The formalism advanced here can be used 
for both static and dynamic thin shells. In the next section we will 
make full use of the formalism when we apply it first to find the 
pressure necessary to maintain a charged thin shell in static 
equilibrium in a vacuum background interior, black hole or otherwise, 
and second to the study of the collapse of a charged thin shell into a 
vacuum background interior, black hole or otherwise, in Lovelock 
gravity coupled to Maxwell electromagnetism.  Since the three sectors 
that enter the problem are the gravitational background, the 
electrodynamic interaction, and the matter that constitutes the shell, 
we have to set up the metrics for the interior and exterior to the 
shell, we have to give the form of vector potential field, and we have 
also to give the energy momentum tensor. 

\vskip 0.2cm 
\noindent {The gravitational Hamiltonian in spherically symmetric 
spacetimes:} 
\vskip 0.1cm 
\noindent 
We assume now the shell traces its spacetime 
trajectory in a static spherically 
symmetric background. Thus the generic ansatz both for the 
exterior and interior to the shell can be written as 
\begin{equation} 
ds^2=-N^2(r)f^2(r)dt^2+\frac{dr^2}{f^2(r)}+r^2d\Omega^2_{d-2}\,, 
\label{staticsphericallysymmetricansatz} 
\end{equation} 
where Schwarzschild type coordinates 
$\{t,\,r,\,\theta^1,\,\theta^2,\cdots,\,\theta^{(d-2)}\}$ have been 
chosen, $N(r)$ is the lapse function exclusively dependent on the radial 
coordinate, 
$f(r)$ is the metric function, dependent on $r$ only, and 
$d\Omega^2_{d-2}$ is the line element of the $d-2$-dimensional unitary 
sphere, written explicitly as 
$d\Omega^2_{d-2}=(d\theta^1)^2+\sin(\theta^1)^2(d\theta^2)^2+ 
\sin(\theta^1)^2\sin(\theta^2)^2(d\theta^3)^2+\cdots+ 
\prod_{i=1}^{d-3}\sin(\theta^i)^2\,(d\theta^{(d-2)})^2$. 
With the ansatz (\ref{staticsphericallysymmetricansatz}), the 
first 
gravitational constraint (\ref{normalgravitationalhamiltonian}) is written 
as 
\begin{eqnarray} 
\mathcal{H}^{(g)}_{\bot}&=& -\kappa 
\frac{(d-2)!}{r^{d-2}}\sqrt{g} \frac{d}{dr}\left\{ r^{d-1}\sum_p 
(d-2p)\,\alpha_p \left( \frac{1-f^2}{r^2} \right)^p\right\}\,, 
\label{gravitationalconstraint} 
\end{eqnarray} 
and the other gravitational constraints 
(\ref{surfacegravitationalhamiltonian}) yield the equation 
\begin{eqnarray} 
\mathcal{H}^{(g)}_{i}&=&\kappa\sqrt{\Omega} 
(d-2)!\sum_{p=0}^{k}\frac{p!(d-2p)!\alpha_p}{2^{p+1}(d-2p-1)!} 
\sum_{s=0}^{p-1}2^s D_{s(p)}f^{2(s-p)}(1-f^2)^s\times \nonumber \\ 
&&\times\frac{d}{dr}\left(r^{d-2p-1}(\alpha_p\dot{r})^{2p-2s-1}\right)\,, 
\label{gravitationalconstraintother} 
\end{eqnarray} 
where $\Omega$ is the angular part of the determinant $g$ 
of the intrinsic metric $g_{ij}$ of the hypersurface $\Sigma_t$, with 
$g=g_{rr}r^{d-2}\Omega$, and we use the fact that the metric is 
diagonal. The function $D_{s(p)}$ is defined in Eq. 
(\ref{dfunction}). 

\vskip 0.2cm 
\noindent {The electrodynamic Hamiltonian in spherically symmetric 
spacetimes:} 
\vskip 0.1cm 
\noindent 
For a static system with spherical symmetry the electromagnetic field is 
purely electric and radial for an outside observer at rest.  In this 
case, the vector potential has only one non-zero component, which 
depends exclusively on the radial coordinate. 
\begin{equation} 
A_t=A(r)\,. 
\label{staticsphericallysymmetricelectricfield} 
\end{equation} 
Given the symmetries and the fact that we are working on 
a static background, the totally spatial components of the Maxwell 
tensor are null, i.e., $F_{ij}=0,\,\, i,j=r, 
\theta^1,\cdots,\theta^{(d-2)}$. 
The only 
non-vanishing component of the Maxwell tensor is $F_{tr}=-\partial_r 
A_t (r)$, with also 
$F_{rt}=\partial_r 
A_t (r)$. Thus, electrically charged, 
static, spherically symmetric vacuum solutions imply 
\begin{eqnarray} 
F_{ij}&=& 0 \label{1}\,,\\ 
p^i&=& \left( 0,p^r,0,\ldots,0 \right) \label{2}\,,\\ 
\dot{A}_i&=& 0\;\; {\rm and} \;\;  \dot{p}^i=0 \label{3}\,, 
\label{emfields} 
\end{eqnarray} 
where Eq. (\ref{1}) means there is no magnetic field, 
Eq. (\ref{2}) means the electric field is spherically symmetric, 
and Eq. (\ref{3}) means the 
field is static. From equations 
(\ref{emnormalconstraint})-(\ref{emconstraint}), 
the electrodynamic constraints, associated with the Lagrange 
multipliers $N^{\bot}$ and $\varphi$, are, respectively, 
\begin{eqnarray} 
\mathcal{H}_\bot^{(e)}&=& 
\frac{1}{2\,\sqrt{g}} \Omega_{d-2}\,p^rp_r 
=\frac{1}{2\,\sqrt{g}}\Omega_{d-2}\,(p^r)^2f^{-2}(r)\,, 
\label{electromagneticconstraint1}\\ 
\mathcal{H}^{(e)}_i &=& 0\,, 
\label{electromagneticconstraint3}\\ 
E_{\varphi} &=& p^r_{,r}-J^0\,, 
\label{electromagneticconstraint2} 
\end{eqnarray} where $f(r)$ comes from 
(\ref{staticsphericallysymmetricansatz}). 

\vskip 0.2cm 
\noindent {The thin shell Hamiltonian in spherically symmetric 
spacetimes:} 
\vskip 0.1cm 
\noindent 
We want to spell out completely the matter constraints, namely, 
$\mathcal{H}_\bot^{(m)}=\sqrt{g}\,T_{\bot\bot}$ and 
$\mathcal{H}_i^{(m)}= 2\,\sqrt{g}\,T_{\bot i}^{(m)}$, for 
a charged thin shell in a charged vacuum 
background geometry. More precisely we 
want to develop a Hamiltonian framework for this situation 
and set up a natural junction which arises in this 
formalism. For this to be made we have first to describe the 
geometrical set-up in order to be able to write the projected 
components of the energy-momentum tensor, namely, $T_{\bot\bot}$ and 
$T_{\bot i}$. 

\begin{figure}[ht] 
\includegraphics*[height=6 cm]{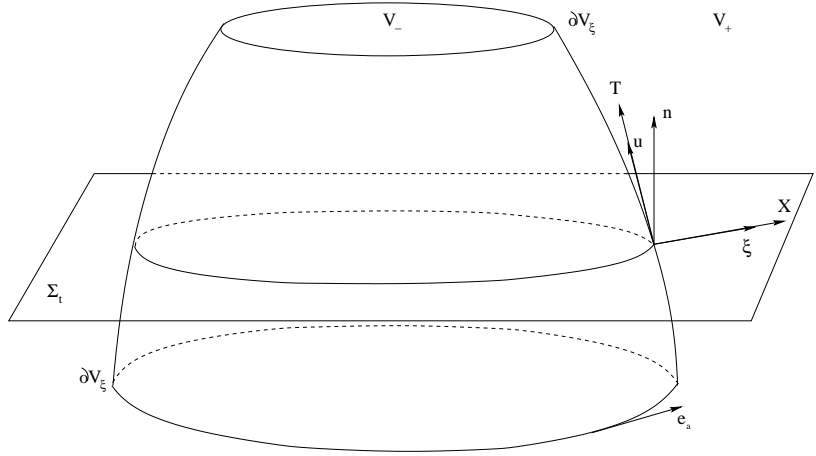} 
\caption{\label{g1} The shell trajectory manifold 
$\partial V_\xi$ in spacetime is depicted. 
$V_-$ and  $V_+$ are the spacetime 
regions interior and exterior to $\partial V_\xi$, 
respectively, and $\Sigma_t$ is the hypersurface 
intersecting $\partial V_\xi$. 
Also shown are the normal vector $n$ (normal 
to $\Sigma_t$), the velocity vector $u$ (tangent to $\partial V_\xi$), 
the vector $\xi$ (the spacelike normal to $\partial V_\xi$), and 
the vector $e_a$ (the generic tangent vector to $\partial V_\xi$). 
The vectors indicating the origin and direction of the adapted 
coordinate system $\{ T,X \}$ are depicted together with the $u$ and 
$\xi$ vectors.} 
\label{shellmanifold} 
\end{figure}

First we state the nomenclature for the thin shell and for the 
interior and exterior spacetimes.  The thin shell is a 
$d-2$-dimensional spacelike surface, which evolves in time, see Figure 
\ref{shellmanifold}.  This time evolution of the thin shell can be 
represented by a timelike hypersurface $\partial V_\xi$, a boundary 
surface, which divides spacetime into two regions, the interior, 
denoted by $V^{(-)}$, and the exterior, denoted by $V^{(+)}$. At each 
point on the boundary surface there exists a unit space-like vector 
$\xi$, with components 
$\xi^\mu$, normal to $\partial V_\xi$, and pointing from the interior 
$V^{(-)}$ to the exterior $V^{(+)}$.  In $\partial V_\xi$ there is a 
set of intrinsic spacetime coordinates $\rho^a$, where $a$ runs from 
$0$ to $d-2$. In the regions $V^{(-)}$ and $V^{(+)}$ independent 
coordinates are introduced, $x^\mu_-$ and $x^\mu_+$ 
respectively (where $\mu$ runs from 
$0$ to $d-1$), and so the parametric equations for $\partial V_\xi$ in 
these coordinates are $x^\mu_-(\rho^a)$ and 
$x^\mu_+(\rho^a)$. Thus, $\partial V_\xi$ has tangential vectors 
$e_a$ with components given by $e^\mu_a=\frac{\partial 
x^\mu}{\partial \rho^a}$.  Since $\partial V_\xi$ represents the 
spacetime evolution of the $d-2$-dimensional thin shell, the 
$d$-velocity $u$, with components $u^\mu$, 
of the matter of the shell is tangential to 
$\partial V_\xi$, with $u$ being orthogonal to $\xi$, and 
vanishing outside the shell.  One can also consider an intrinsic 
velocity vector $\bar{u}$, with components 
$\bar{u}^a$, related to the velocity $u$ through the relation 
$u^\mu=e^\mu_a \bar{u}^a$, where $\bar{u}$ is built by considering 
the evolution of matter in the shell using coordinates 
$(\rho^1,\cdots,\rho^{d-2})$ intrinsic to the shell, and the intrinsic 
time $\rho^0$. 

Working with the spherically symmetric metric 
(\ref{staticsphericallysymmetricansatz}), we choose the 
intrinsic coordinates of the thin shell surface $\partial V_\xi$ as 
the proper time $\tau$, 
and the spherical angles of the Schwarzschild-like 
coordinates, 
so that $\rho^a=(\tau,\theta^1,\ldots,\theta^{d-2})$. The motion 
of the shell is governed by the radial function of the proper time 
$r=R(\tau)$. The derivative with respect to $\tau$ will henceforth be 
denoted by a dot, e. g. $\frac{d}{d\tau}R(\tau)\equiv \dot{R}(\tau)$. 
The line element for $\partial V_\xi$ in these coordinates is 
\begin{equation} 
ds^2_{\Sigma} = -d\tau^2 + 
R^2(\tau)d\Omega_{d-2}^{2}\,. 
\label{shelllineelement} 
\end{equation} 
We want to study the spacetime evolution of a thin shell in a vacuum 
background with metrics of the type given in 
(\ref{staticsphericallysymmetricansatz}). Thus, the interior 
$V^{(-)}$ and exterior $V^{(+)}$ spacetimes have their general line 
element given by 
\begin{eqnarray} 
ds^2_{-}&=&-f^2_-(r)dt^2_-+f^{-2}_-(r)dr^2+r^2d\Omega^2_{d-2}\,, 
\label{inlineelement}\\ 
ds^2_{+}&=&-f^2_+(r)dt^2_++f^{-2}_+(r)dr^2+r^2d\Omega^2_{d-2}\,. 
\label{outlineelement} 
\end{eqnarray} 
The radial and angular coordinates match continuously 
at the shell, but the time coordinates do not in general. The radial 
and angular coordinates match on account of the fact that the surface 
area of the shell is independent of the coordinates used to determine 
it, and it should be the same on both sides, as we are working in the 
limit of infinitesimal thickness, the thin shell limit. As the area 
of the shell is a function of the radial position of the shell 
$R(\tau)$, there should not be a different radius for the shell inside 
and outside, as we would be attributing two different areas to the 
shell. Also, spherical symmetry allows us to induce coordinate 
charts on the manifold $\partial V_\xi$ merely by restricting the charts 
of $V^{(-)}$ and $V^{(+)}$ to their boundary on $\partial V_\xi$. As, due 
to symmetry, the radial and angular coordinates on both sides of the 
shell are the same, their restriction to the shell should be the same, 
and hence match perfectly. However, the restriction of the inside line 
element (\ref{inlineelement}) to the surface of the shell must match 
the restriction of the outside line element (\ref{outlineelement}), 
all having the general form (\ref{shelllineelement}) on the shell. 
In these coordinates the vectors $u$ and $\xi$ defined 
above (see also Figure \ref{shellmanifold}) have the following 
components, 
\begin{eqnarray} 
u^\mu &=& \left( \frac{\gamma}{f^2}, \dot{R}, 0, 
0 \right)\,, \label{d-1velocity}\\ 
\xi^{\mu} &=& \left( 
\frac{\dot{R}}{f^2}, \gamma, 0, 0 \right)\,,\label{spacelikenormal} 
\end{eqnarray} 
where 
\begin{equation} 
\gamma = \sqrt{f^2+\dot{R}^2}\,, 
\label{gamma} 
\end{equation} 
where 
$\gamma_\pm$ is a generalized Lorentz factor. 
These vectors were obtained through the matching of 
(\ref{inlineelement}) and (\ref{outlineelement}) to the surface line 
element (\ref{shelllineelement}), using the relation 
$u^\mu=\frac{dx^\mu}{d\tau}$, where the $t_\pm$ coordinates and 
the radial coordinate are treated as functions of the proper time 
$\tau$.  We have also used $u^\mu u_\mu=-1$, $\xi^\mu 
u_\mu=0$, and $\xi_\mu \xi^\mu=1$ (as $\xi$ is the 
spacelike normal to $\partial V_\xi$). 

The shell matter properties are described by the surface 
energy-momentum tensor $T_{ab}$, which is orthogonal to $\xi$ 
and vanishes outside the hypersurface $\partial V_\xi$.  For an 
observer in the shell, $T_{ab}$ is written in intrinsic coordinates, 
which for a perfect fluid is given by 
\begin{equation} 
T_{ab}=\sigma \bar{u}_a \bar{u}_b + P \, (h_{ab}+ \bar{u}_a \bar{u}_b )\,, 
\end{equation} 
where $\sigma$ is the rest mass surface density, 
$P$ is the surface pressure (or, when negative, the surface 
tension), 
and $h_{ab}\equiv g_{\mu\nu}e^\mu_ae^\nu_b$. 
$T_{ab}$ is confined to the hypersurface $\partial V_\xi$ 
and it satisfies the conservation equation, $T_{a|b}^b=0$, where a $|$ 
denotes covariant differentiation on the hypersurface, 
which implies the explicit equation 
\begin{equation} 
(\sigma \bar{u}^a)_{|a}+P\, \bar{u}^a_{|a}=0\,, 
\label{matterequationofstate} 
\end{equation} 
this last being obtained by multiplying the tensor 
$T_{ab}$ by $\bar{u}^a$. 

To apply the Hamiltonian formalism we should pass to a spacetime 
characterization of the energy-momentum tensor.  In order to do so we 
define the spacetime energy-momentum tensor as 
\begin{equation} T^{\mu\nu}=T^{ab}e_a^\mu\,e_b^\nu\,. 
\label{spacetimeTmunu} 
\end{equation} 
Then, in order to write $T_{\bot\bot}$ and $T_{\bot 
i}$ we must first write the energy-momentum tensor in an adapted 
coordinate system $\{T,X,\theta^1,\cdots,\theta^{d-2}\}$ (see Figure 
\ref{shellmanifold}), where $T$ and $X$ are the time and 
radial adapted coordinates, and 
$\theta^1,\cdots,\theta^{d-2}$ are the usual angular coordinates. 
In more detail, the origin 
of the new adapted coordinate $X$ is at the shell, and the direction 
of any vector in $X$ alone is the same as that of 
the outward spacetime normal to the surface, $\xi$, and 
the direction 
of any vector in $T$ alone is the same as that of 
the tangent vector $u$. Then, 
\begin{equation} 
T^{\mu\nu}=\left\{\sigma u^\mu u^\nu + P \, 
(h^{\mu\nu}+ u^\mu u^\nu )\right\}\delta (X)\,, 
\label{emtensordeltaform} 
\end{equation} 
where $h^{\mu\nu}$ is the intrinsic metric of the 
shell written as a bulk $d$ dimensional spacetime tensor, 
$h^{\mu\nu}=g^{\mu\nu}-\xi^\mu \xi^\nu\,, 
\label{intrinsicmetric} 
$ 
and $u^{\mu}$ is the velocity of the particles of the 
shell written as a bulk $d$ dimensional vector.  When written as a 
bulk tensor, the energy momentum tensor of the thin shell is proportional 
to a delta function. 
Now, $T_{\bot\bot}$ is given by $T^{\mu\nu}n_\mu n_\nu$, which means 
that 
$ 
T^{\mu\nu}n_\mu n_\nu=\left[\sigma u^\mu u^\nu 
n_\mu n_\nu + P \, (h^{\mu\nu}n_\mu n_\nu+ u^\mu u^\nu n_\mu 
n_\nu)\right]\delta (X)\,. 
$ 
Knowing that $n^\mu n_\mu = -1$, we have that 
$(N^\bot)^2 = f^2$. So 
$u_\bot u_\bot = \gamma^2/f^2\,.$ In the same way, we obtain 
$ 
h^{\mu\nu}n_\mu n_\nu = -1-(\dot{R}^2)/(f^2)\,. 
$ 
We arrive thus at 
$ T_{\bot\bot} = \sigma 
\left(\gamma^2/f^2\right)\delta(X)\,. 
\label{embot} 
$ 
It is, however, necessary to write the delta function as a function 
of the radial coordinate $r$. This means we have to return to the 
original Schwarzschild type coordinates of the metrics 
(\ref{inlineelement}) and (\ref{outlineelement}). For that, we write 
$ 
dt = u^t dT + \xi^t dX,\,\, 
dr = u^r dT + \xi^r dX\,. 
\label{diffcoordrelations} 
$ 
As we are interested in the energy-momentum tensor on 
the spacelike hypersurface $\Sigma_t$, we make $dt=0$ and obtain the 
differential relation 
$ 
dr/dX = \xi^r-(u^r/u^t)\xi^t = 
f^2/\gamma\,. 
\label{derivativecoordinates} 
$ 
Taking into consideration the relation 
$ 
\delta(f(x)) = \Sigma_i 
\delta(x-x_i)/\left|f'(x_i)\right|, 
\label{deltalaw} 
$ where $x_i$ are the zeros of a general function 
$f(x)$, and $'$ denotes differentiation with respect to the argument 
of the function, we have 
$ 
\delta(X) = \left(f^2/\gamma\right)\delta[r-R(\tau)]\,. 
\label{deltatransformation} 
$ 
Plugging one equation into 
the other we get 
\begin{equation} 
T_{\bot\bot} = \sigma \gamma \delta[r-R(\tau)]\,, 
\label{bot} 
\end{equation} 
where $\sigma$ is the energy surface density on the 
shell, $\gamma$ is given in (\ref{gamma}), and $R(\tau)$ is the shell 
radial function. 

For the other components of the matter tensor, $T_{\bot 
i}$, we have first to define $z^\mu_i = \frac{\partial 
x^\mu}{\partial y^i}$ as the projection vectors onto the 
hypersurface $\Sigma_t$, where $y^i$ are the intrinsic coordinates of 
$\Sigma_t$. Then, writing the intrinsic metric of $\partial V_\xi$ as 
$h^{\mu\nu}= g^{\mu\nu}-\xi^\mu \xi^\nu$, and knowing 
that $u^\mu \xi_\mu=0$, we arrive at 
\begin{equation} T_{\bot i} = \frac{f^2}{\gamma}\left[(\sigma + 
P)(-N^\bot f^{-2} \gamma) u_i + P u_i \right]\delta[r-R(\tau)]\,, 
\label{emboti} 
\end{equation} where $u_i$ are the components of the velocity vector 
projected onto $\Sigma_t$. 
It is then possible to write completely the matter constraints as 
\begin{equation} 
\mathcal{H}_\bot^{(m)}=\sqrt{g}\,\sigma \gamma 
\delta[r-R(\tau)]\,, 
\label{matterconstraint2} 
\end{equation} and 
\begin{equation} 
\mathcal{H}_i^{(m)} = 
2\,\sqrt{g}\,\frac{f^2}{\gamma} \left[(\sigma + P)(-N^\bot f^{-2} 
\gamma) u_i + P u_i \right]\delta[r-R(\tau)]\,. 
\end{equation} 

One remark is due, regarding the foliation of spacetime. From 
the way we have sliced the spacetime, 
by looking at the ADM metric (\ref{ADMmetric}) and the fact 
that we are working with spherical symmetry and static background 
(cf. (\ref{staticsphericallysymmetricansatz})), we see that the shift 
Lagrange multipliers $N^i$ are equal to zero. Therefore, every 
constraint with components in the $\Sigma_t$ hypersurface 
$\mathcal{H}_i$ will not be relevant to our discussion. 
In the following we will need neither 
$T_{\bot i}$ nor $\mathcal{H}_i^{(m)}$. 

\subsubsection{The equations of motion} 

                   By varying the Hamiltonian action, written with the 
constraints 
spelt out previously, with respect to $g$, 
$N^\bot$, $\varphi$, and $p^r$ the field equations will be, 
respectively 
\begin{eqnarray} 
\frac{dN}{dr}&=& 0 \,, 
\label{chargedfieldequations1}\\ 
-\kappa \frac{(d-2)!}{r^{d-2}} 
\frac{d}{dr}\left( r^{d-1} \left[F(r)+\frac{1}{l^2}\right]^k 
\right)&=& \frac{p^2} {2\, \Omega_{d-2}} + T_{\bot\bot} \,, 
\label{chargedfieldequations2}\\ 
\frac{d}{dr}\left( r^{d-2} p \right)&=& r^{d-2}\,j^0 \,, 
\label{chargedfieldequations3}\\ 
\frac{d\varphi}{dr}+ N\,p &=& 0 \,, 
\label{chargedfieldequations4} 
\end{eqnarray} 
where $F(r)=(1-f^2(r))/r^2$, $N=N^\bot 
(g_{rr})^{\frac12}$, and $p(r)$ is a redefinition of the canonical 
momentum radial component $p^r$ through 
\begin{eqnarray} 
p^r &=& r^{d-2}\frac{\Omega^{\frac12}}{\Omega_{d-2}}p\,, 
\end{eqnarray} 
where $\Omega$ is the angular part of the determinant $g$ 
of the intrinsic metric $g_{ij}$ of the hypersurface $\Sigma_t$, with 
$g=g_{rr}r^{d-2}\Omega$, 
and we have used the fact that the metric is 
diagonal. In the same way we have redefined the current's time 
component 
\begin{eqnarray} 
J^0 &=& 
r^{d-2}\frac{\Omega^{\frac12}}{\Omega_{d-2}}j^0\,. 
\end{eqnarray} 
We also have used (\ref{alphaconstants}) to write 
\begin{eqnarray} \sum_p (d-2p)\, \alpha_p 
\left(\frac{1-f^2}{r^2}\right)^p&=& \left( F+\frac{1}{l^2} 
\right)^k\,. 
\label{binomialidentity} 
\end{eqnarray} One should note that varying with respect to $N^\bot$ 
and $\varphi$ is requiring that the constraints be made equal to zero, 
because both are Lagrange multipliers. That is, in the case of 
$N^\bot$, the variation implies $\mathcal{H}_\bot=0$, and varying with 
respect to $\varphi$ implies $E_\varphi=0$. 

Equations 
(\ref{chargedfieldequations1})-(\ref{chargedfieldequations4}) are 
valid for any thin spherically symmetric charged shell 
in Lovelock gravity coupled to Maxwell electromagnetism. One has only to 
give 
the thin shell delta-function energy-momentum tensor and integrate the 
equations to find the solutions.  Note that the Eq. 
(\ref{chargedfieldequations2}) is valid only for a delta function 
energy-momentum tensor. Indeed this equation when integrated gives the 
equation of motion of the shell itself, but if one has a continuous 
distribution of matter, instead of a thin shell, then within the 
matter there exists time dependence in the metric function, which 
should be taken into account. In the case of the thin shell we were 
able to ignore this time dependence, since inside and outside the 
shell it is possible to set up static backgrounds. 

\section{Thin shell solutions in a vacuum background with 
spherical symmetry} 
\label{solutions} 

\subsection{Spherically symmetric vacuum solutions: 
solutions with black hole character, with naked singularity 
character, and empty solutions} 

The vacuum solutions are 
obtained when we consider the 
matter action equal to zero, $T_{\bot\bot}=0$, and when there is no 
vector current, $J^\mu=0$. So, after integrating the field equations 
above, one obtains the vacuum solutions \cite{CTZ} 
\begin{eqnarray} 
N&=&N_{\infty}=1\,, 
\label{chargedsolution1}\\ 
f^2(r)&=& 1+\frac{r^2}{l^2}-\chi \, 
g_k(r)\,,\label{chargedsolution4}\\ p(r)&=& \epsilon 
\frac{Q}{r^{(d-2)}}\,,\label{chargedsolution2}\\ 
\varphi(r)&=& 
\varphi_\infty + \frac{\epsilon}{(d-3)}\frac{Q}{r^{(d-3)}}\,. 
\label{chargedsolution3} 
\end{eqnarray} 
The function $g_k(r)$ is defined by 
\begin{equation} 
g_k(r)=\left( 
\frac{2G_kM+\delta_{d-2k,1}}{r^{d-2k-1}} -\frac{\epsilon\, 
G_k}{(d-3)}\frac{Q^2}{r^{2(d-k-2)}} \right)^{1/k}\,, 
\label{gfunction} 
\end{equation} 
where the integration constants $M$ and $Q$ are the mass 
and electric charge of the solutions, 
and where we have used Eq. (\ref{kappa}) for the definition 
of $\kappa$ in terms of $G_k$. From 
Eqs. (\ref{chargedsolution1})-(\ref{gfunction}), in particular 
from  Eqs. (\ref{chargedsolution2}) and (\ref{gfunction}), 
one notes that the vacuum solutions 
yield a natural division of the Lovelock theory into two branches, 
namely $d-2k-1>0$ and $d-2k-1=0$, with $d$ being the 
dimension of the spacetime and $k$ the parameter that 
gives the highest power of the curvature in the Lagrangian. 
The branch $d-2k-1>0$  embodies general relativity when 
$k=1$ (any $d$), Born-Infeld when $k=[\frac{d-1}{2}]$, 
and other generic cases. The branch $d-2k-1=0$ 
embodies Chern-Simons type theories alone. 
Worth mentioning now 
is the fact that in the $d-2k-1>0$ branch 
the empty vacuum solutions 
have $M=0$ and $Q=0$, 
whereas in the $d-2k-1=0$ branch 
the empty vacuum solutions 
have $M=-(2G_k)^{-1}$ and $Q=0$. 
In the solutions, there 
appears an additional unusual parameter 
$\chi$. The parameter 
$\chi$ is the spacetime character,  given by 
\begin{equation} 
\chi=(\pm 1)^{k+1}\,. 
\label{chi} 
\end{equation} 
This means that the vacuum solutions may 
have a black hole character or a naked singularity character. 
For 
$\chi=1$ the solutions, being of the type found in general relativity, 
have a black hole character, since for a correct choice of the 
parameters, such as mass and charge, the solution is a black hole 
solution, although, of course, for other choices of parameters it can 
be an extremal black hole or a naked singularity.  For $\chi=-1$ the 
solutions, being of a new type not found in general relativity, have a 
naked singularity character, as there is no possible choice of 
the parameters that gives a black hole solution, the full vacuum 
solution is always singular without horizons. 
Note also that Eq.  (\ref{chi}) implies that if $k$ is odd (such as in 
general relativity, where $k=1$) the character $\chi=1$ only, whereas 
if $k$ is even (such as in general relativity with a Gauss-Bonnet 
term, where $k=2$) the character $\chi$ can have both values $\pm1$. 
Note also that in the $d-2k-1=0$ 
Chern-Simons theory, the black 
hole vacuum spacetime, i.e., the vacuum of the $\chi=1$ character 
black hole solution, given by $M=0$ and 
$Q=0$, is different from the usual AdS spacetime, and there is 
a mass gap between these spacetimes, with the latter being obtained for 
$M=-(2G_k)^{-1}$, as mentioned above. 
Equations (\ref{chargedsolution1})-(\ref{chi}) provide the 
electrically charged solutions of the Lovelock theory 
defined by the parameters $d$ and $k$ 
coupled to Maxwell electromagnetism, 
with $d\geq4$. The dimension $d=3$ yields singular 
solutions in the charged sector, and so we do not consider 
this dimension in this work, although in the uncharged 
sector it yields perfectly sensible solutions \cite{BTZ1}. 

Now, given a spacetime solution one should search for horizons and 
singularities. For $\chi=1$ the solution has a black hole 
character and can have horizons and singularities, 
i.e., the solutions represent black holes when the parameters 
are appropriately chosen. 
For $\chi=-1$ the solution has a naked singularity character, 
and should have only singularities with no horizons, 
independently of the choice of the other parameters. 
First we locate the horizons for the $\chi=1$ solutions, 
and then we locate the singularities for both $\chi=\pm1$. 
For $\chi=1$, the 
zeros of $f(r)$, in the coordinates used, 
give the horizons.  Upon close scrutiny, the properties 
of these solutions have many similarities with the $d$ dimensional 
Reissner-N\"{o}rdstrom-AdS black holes in pure general relativity. 
Moreover, this set of black hole solutions reduces to the $d$ dimensional 
Reissner-N\"{o}rdstrom-AdS black holes for $k=1$, and the charged 
Born-Infeld and charged Chern-Simons black hole solutions are 
recovered for $d=2k+2$ and $d=2k+1$, respectively \cite{BTZ1}. 
For generic values 
of $d$ and $k$, in analogy with the Reissner-Nordstr\"om geometry, the 
black hole solutions possess, in general, two horizons located at the 
roots of $f^2(r)$, one is the event horizon $r_+$, and the other 
the Cauchy horizon $r_-$, with $r_-<r_+$. When the two horizons merge the 
black hole is extremal as usual.  If the solution is overcharged one 
has a naked singularity for some value of $Q$ in terms of $M$ 
\cite{CTZ}.  Now one finds the location of the singularities, 
which can be treated together for both characters $\chi=\pm1$. 
The scalar curvature is 
\begin{equation} R=\frac{1}{r^{d-2}}\frac{d^2}{dr^2}\left[ r^{d-2} 
\left( g_k(r)-\frac{r^2}{l^2} \right) \right]\,. 
\label{scalarcurvatureknot1} 
\end{equation} 
For $k=1$, general relativity with its Reissner-N\"{o}rdstrom-AdS 
black holes, one finds that $R=0$ as it should, 
thus in this case the singularities are located when the Kretschmann 
scalar, 
$R_{\mu\nu\rho\sigma}\,R^{\mu\nu\rho\sigma}$, blows up, 
which is at $r=0$.  For $k>1$ the Ricci scalar is no more zero, 
$R\neq0$.  In this case, Eq. 
(\ref{scalarcurvatureknot1}) has a singularity at the zero of the 
function $g_k(r)$, and is due 
to the existence of an electric field. 
This is a real timelike singularity located at $r_{e}$ 
\begin{equation} 
r_{e}=\left[ 
\frac{\epsilon}{2(d-3)}\frac{Q^2}{(M+(2\,G_k)^{-1} \delta_{d-2k,1})} 
\right]^{1/(d-3)},\quad k>1\,. 
\label{electricradius} 
\end{equation} 
This singularity can be reached in a finite proper time interval. 
However, an external observer is protected from it because both 
horizons cover it, $r_{e} < r_- < r_+$.  Moreover, for even $k$, 
regions where $r<r_{e}$ have a metric with complex functions, so 
there is no solution in this region. For odd $k$, the metric can be 
defined in the region $0<r<r_{e}$, where $r=0$ is also a 
spacetime timelike singularity. This solution has no horizons, and is 
defined 
between two timelike singularities and so is of no interest to 
us. For all purposes the solutions of interest, for $k>1$, 
are defined in the interval 
$r_{e}<r<\infty$, and for $k=1$ (general relativity) 
in the interval $0<r<\infty$. 
The corresponding Carter-Penrose diagrams can be 
easily sketched. 

\subsection{Spherically symmetric 
thin shell solutions through the Hamiltonian formalism} 
\label{section3} 

\subsubsection{Shell dynamics} 
\label{shelldynamics} 

\vskip 0.3cm 
{\noindent (a) The two master equations} 
\vskip 0.1cm 

We now want to find the equations governing the motion of a thin shell 
in a vacuum background of a given Lovelock theory specified by the 
dimension $d$ of the spacetime and the parameter $k$, which gives the 
highest power in the curvature terms in the Lagrangian of the theory. 
In order to obtain shell solutions in a vacuum background, as opposed 
to pure vacuum solutions, we study the complete field equations, 
Eqs. (\ref{chargedfieldequations1})-(\ref{chargedfieldequations4}), 
and take into consideration the matter term $\mathcal{H}^{(m)}$ and 
the electric current $J^\mu$ in those equations.  We know that inside 
the shell the spacetime solution, with mass $M_-$ say, is 
obtained by integrating the constraint $\mathcal{H}_{\bot}=0$, i.e., 
the Eq. (\ref{chargedfieldequations2}) with $T_{\bot\bot}=0$, 
with the vacuum solution given in (\ref{chargedsolution4}).  In the 
same way, outside the shell the spacetime vacuum solution, with mass 
$M_+$ say, is obtained by integrating the constraint 
$\mathcal{H}_{\bot}=0$, i.e., the Eq. 
(\ref{chargedfieldequations2}) again with $T_{\bot\bot}=0$, with the 
vacuum solution given in (\ref{chargedsolution4}). It remains now to 
integrate the constraint $\mathcal{H}_{\bot}=0$ in the neighborhood of 
the shell, from $R-\epsilon$ to $R+\epsilon$, where $\epsilon$ is the 
infinitesimal thickness of the shell, with $\epsilon\rightarrow0$. 
That is, we have to impose $\int_{R-\epsilon}^{R+\epsilon} dr \, 
\mathcal{H}_{\bot}=0$.  This integration should be performed in the 
asymptotic region, outside of the black hole, since there 
the normal to the $t={\rm constant}$ hypersurfaces is timelike, and 
thus the Hamiltonian formalism is directly applicable.  In addition, one 
has to 
fix from the start the same value of the character 
$\chi$ on both sides of the shell 
(one has to choose either $\chi=1$ or $\chi=-1$), i.e., one has to 
choose which of the two types of spacetime one is working with.  Using 
then Eq. (\ref{chargedfieldequations2}) we obtain 
\begin{equation} 
\int_{R-\epsilon}^{R+\epsilon} dr \left(-\kappa 
\frac{(d-2)!}{r^{d-2}} \frac{d}{dr}\left\{ r^{d-1}\left[ 
F+\frac{1}{l^2} \right]^k\right\} -\frac{p^2}{2\,\Omega_{d-2}} - 
T_{\bot\bot}\right) =0 \,, 
\label{integrationofequations1} 
\end{equation} 
where $T_{\bot\bot}$ is given in (\ref{bot}). 
Knowing (\ref{alphaconstants}) and (\ref{kappa}), using 
(\ref{binomialidentity}), and defining the energy-density content of 
the shell $m\equiv\sigma\,\Omega_{d-2}R^{d-2}$, the integration yields 
\begin{eqnarray} 
\frac12 m \,(\gamma_+ +\gamma_-)&=& 
\left( M_+ - M_- \right) - \frac{\epsilon (Q_+^2 - 
Q_-^2)}{2 (d-3)}\frac{1}{R^{d-3}} \,, 
\label{shellequation2} 
\end{eqnarray} 
where following Eq. (\ref{gamma}) 
\begin{equation} 
\gamma_+\equiv\sqrt{f^2_+ + \dot{R}^2}\; 
{\rm and}\; \gamma_-\equiv\sqrt{f^2_- + \dot{R}^2}\,, 
\label{gammaplusminus} 
\end{equation} 
$\gamma_\pm$ being the generalized Lorentz factor, and where the 
different indices $+$ and $-$ stem from the fact that the integration 
is made both in the $V^{(-)}$ (from $R-\epsilon$ to $R$) and $V^{(+)}$ 
(from $R$ to $R+\epsilon$) spacetimes. Note that when integrating 
Eq.  (\ref{integrationofequations1}), we have used the fact 
that the radial electric field is zero inside a uniformly charged 
sphere, due to Gauss' theorem, and that outside it is 
$(Q_+-Q_-)/(r^2+\epsilon^2)$, where $\epsilon$ denotes the distance from 
the 
surface on the outside to the point of measurement, $r$ denotes the 
radius of the sphere, and $f^{-2}(r)$ is continuous in the domain of 
integration. When there are black holes this domain should 
be exterior to horizon of the 
black hole solution, where the solutions are static. 
Then, the integration  in $r$ of the 
$\mathcal{H}_\bot^{(e)}$ 
between the limits $R-\epsilon$ and $R+\epsilon$ is, in 
the limit of $\epsilon\rightarrow 0$, equal to zero. Thus, the 
electric field does not contribute to the shell equation 
through the radial integration of the Hamiltonian constraint.  It is 
sometimes useful to write Eq. (\ref{shellequation2}) in another way, 
namely to multiply both sides by $\gamma_- - \gamma_ +$. This yields 
the following equivalent equation 
\begin{equation} 
\,m\,\chi\,\left(\,g_{k+}(R)-\,g_{k-}(R)\right)= 
\left[g_{k_+}(R)^k-g_{k-}(R)^k \right]\,R^{d-2k-1}\,(\gamma_- - 
\gamma_+)\,. 
\label{equation3} 
\end{equation} 
Eq. (\ref{shellequation2}), and its equivalent (\ref{equation3}), 
were deduced 
for the asymptotic region, where the vacuum spacetime is static, 
but in the case a black hole is present or being formed, 
with some care, those equations 
can be extended, by continuity, to the interior black hole region. 

We proceed our study by analyzing now the electrodynamic constraint 
(\ref{chargedfieldequations3}).  Inside the shell the 
electric field solution, $Q_-/r^{d-2}$, is obtained by integrating the 
constraint $E_\varphi=0$, i.e., Eq. (\ref{chargedfieldequations3}) 
with $j^0=0$, yielding the vacuum solution 
(\ref{chargedsolution2}). In the same way, outside the shell 
the electric field solution, $Q_+/r^{d-2}$, is obtained by 
integrating the constraint $E_\varphi=0$, i.e., Eq. 
(\ref{chargedfieldequations3}) again with $j^0=0$, yielding the vacuum 
solution (\ref{chargedsolution2}). It remains now to integrate the 
constraint $E_\varphi$ in the neighborhood of the shell, from 
$R-\epsilon$ to $R+\epsilon$, where $\epsilon$ is the infinitesimal 
thickness of the shell, with $\epsilon\rightarrow0$.  That is, we have 
to impose $\int_{R-\epsilon}^{R+\epsilon} dr \, E_\varphi=0$. Using 
then Eq. (\ref{chargedfieldequations3}), we obtain 
\begin{equation} 
\int_{R-\epsilon}^{R+\epsilon} dr \, 
\frac{d}{dr}\,(r^{d-2}p)= \int_{R-\epsilon}^{R+\epsilon} dr \, r^{d-2} 
j^0\,, 
\label{integrationofequations2} 
\end{equation} 
where $j^0$ is the time component of the vector 
current, and $j^0=\sigma_e\delta(r-R)u^0$, where $\sigma_e$ is the 
surface charge density.  Defining the charge of the shell as 
$q\equiv\sigma_e\Omega_{d-2}R^{d-2}$ yields 
\begin{eqnarray} Q_+-Q_-=q \,. 
\label{gausslaw} 
\end{eqnarray} 
Relation (\ref{gausslaw}) is Gauss' law, and is a 
trivial consequence of the conservation of charge. 

The two master equations are then Eq. (\ref{shellequation2}) (or 
its equivalent (\ref{equation3})) and Eq. (\ref{gausslaw}).  The 
former can be seen as a dynamic equation for $\dot{R}^2$, the latter 
is a static equation that simply expresses the conservation of 
charge. Thus in the rest of our study we concentrate on Eq. 
(\ref{shellequation2}) (or equivalently on  Eq. (\ref{equation3})). 

\vskip 0.3cm 
{\noindent (b) Rewriting the equations to simplify the analysis} 
\vskip 0.1cm 

Equation (\ref{shellequation2}) (or Eq. (\ref{equation3})) 
was derived for the asymptotic flat region, 
thus if there is a black hole present it was derived for the 
region exterior to the event horizon. We should now transform it in order 
to have it written in a more workable manner. For that 
we have to square Eq. (\ref{shellequation2}) appropriately. 
To simplify the notation let us define the 
right hand side of Eq. (\ref{shellequation2}) as $E$, i.e., 
\begin{equation} 
E\equiv \left( M_+ - M_- \right)- \frac{\epsilon 
(Q_+^2 - Q_-^2)}{2 (d-3)}\frac{1}{R^{d-3}}\,. 
\label{definition1} 
\end{equation} 
Note that  from (\ref{shellequation2}) one sees 
that this quantity $E$ is positive for positive shell mass $m$ 
(i.e, positive energy density $\sigma>0$), a condition we always assume 
throughout.  Now we obtain from Eq. (\ref{shellequation2}) a set of 
two equations which give the dynamics of the shell in detail.  First, 
squaring Eq. (\ref{shellequation2}) and using the definition 
(\ref{definition1}), we get for $\dot{R}^2$ the following expression 
\begin{eqnarray} 
\dot{R}^2 &=& \left[ \frac{E}{m}- 
\frac{m}{4\,E}\left( f^2_+ - f^2_- \right) \right]^2 - f^2_- \,, 
\label{squaredspeedequation1} 
\end{eqnarray} 
or, alternatively 
\begin{eqnarray} 
\dot{R}^2 &=& \left[ \frac{E}{m}+\frac{m}{4\,E} \left( f^2_+ - f^2_- 
\right) 
\right]^2 - f^2_+\,. 
\label{squaredspeedequation2} 
\end{eqnarray} 
Second, there is one last step needed for the entire 
set of relevant equations, in the 
asymptotic region, to be written down. This has to do with the 
fact that squaring Eq. (\ref{shellequation2}) yields 
Eq. 
(\ref{squaredspeedequation1}) 
(or, alternatively, 
(\ref{squaredspeedequation2})) 
making the solutions of (\ref{shellequation2}) 
also solutions of (\ref{squaredspeedequation1}) 
(or, alternatively, 
(\ref{squaredspeedequation2})) 
but not all solutions of the 
(\ref{squaredspeedequation1}) (or, alternatively, 
(\ref{squaredspeedequation2})) 
are solutions of (\ref{shellequation2}). 
So, a sufficient condition is found by putting back 
Eqs. 
(\ref{squaredspeedequation1}) and (\ref{squaredspeedequation2}) 
appropriately 
back into Eq. (\ref{shellequation2}). 
One arrives then at 
the following conditions 
\begin{eqnarray} 
f^2_+ - f^2_- \geq -\frac{4}{m^2}\, E^2\,, 
\label{generalcriticalradius1}\\ 
f^2_+ - f^2_- \leq 
\quad\frac{4}{m^2}\, E^2\,. 
\label{generalcriticalradius2} 
\end{eqnarray} 
Equations (\ref{generalcriticalradius1})-(\ref{generalcriticalradius2}) 
define implicitly a constraint radius $r_c$, 
meaning that, in the asymptotic region, a solution for $R(\tau)$ is a 
solution 
of the equation of motion only if it obeys 
(\ref{squaredspeedequation1}) (or, alternatively, 
(\ref{squaredspeedequation2})) and $R(\tau)>r_c$ 
for all $R$. 
We need now to choose which of the two relations, Eqs. 
(\ref{generalcriticalradius1}) and (\ref{generalcriticalradius2}), 
hold for the system in question. 
In order to 
do so, we have to know what signal the 
difference $f^2_+ - f^2_-$ carries. From (\ref{chargedsolution4}) one 
sees that the most general expression is 
$ 
f^2_+ - f^2_- = \chi\left( g_{k-}(r) - g_{k+}(r)\right)\,, 
\label{diffofffunctions} 
$ 
where $g_k(r)$ comes from (\ref{gfunction}).  The 
signal on right hand side of this equation depends on the 
difference $\chi\left( g_{k-}(r) - g_{k+}(r)\right)$ being positive or 
negative, and this determines which relation in 
(\ref{generalcriticalradius1})-(\ref{generalcriticalradius2}) 
is the relevant one to use.  For 
example, in the case where the interior is flat, $g_{k-}(r)=0$, the 
right hand side is $- \chi g_{k+}(r)$.  In this case when 
$-\chi g_{k+}(r)<0$ the relation 
in (\ref{generalcriticalradius1}) 
is the  relevant one, 
when 
$- \chi 
g_{k+}(r)>0$ the relation in (\ref{generalcriticalradius2}) is 
the appropriate one.  (Note that for pure vacuum, no shell, $- \chi 
g_{k+}(r)>0$, represents a solution with 
naked singularity character, while $- \chi 
g_{k+}(r)<0$ represents a solution with black hole character). 
In brief, Eq. (\ref{generalcriticalradius1}) defines a constraint 
radius $r_c$ which is of use in the asymptotically exterior region. 
So, in the asymptotically exterior region, 
Eqs. 
(\ref{squaredspeedequation1}) (or (\ref{squaredspeedequation2})) 
and (\ref{generalcriticalradius1})-(\ref{generalcriticalradius2}) 
recovers the same solutions as those of the shell equation 
(\ref{shellequation2}), with 
(\ref{generalcriticalradius1})-(\ref{generalcriticalradius2}) being a 
constraint equation.  With this set of equations, containing no square 
roots, one can analyze the spacetime evolution of the shell, in the 
asymptotic region, outside the black hole. 

It is also useful to derive the acceleration of the shell 
radius. Differentiating 
Eqs. (\ref{squaredspeedequation1})-(\ref{squaredspeedequation2}) with 
respect to the proper time of the shell $\tau$, we get 
\begin{eqnarray} 
m\ddot{R}&=& \frac{(Q^2_+ - Q^2_-)\gamma_+ \gamma_- 
m}{2\,E\, \,R^{d-2}}+(d-2) R^{d-3} \Omega_{d-2} P\gamma_+\gamma_- 
\nonumber \\ &-& \frac{m^2}{4\,E}\times \left( \gamma_- 
\frac{df^2_+}{dR} + \gamma_+ \frac{df^2_-}{dR} \right)\,, 
\label{accelerationequation} 
\end{eqnarray} 
where we have used the conservation equation 
(\ref{matterequationofstate}) to arrive at the result 
$\frac{dm}{d\tau}=-(d-2)R^{d-3}P\dot{R}$, which was used in 
(\ref{accelerationequation}). 

\subsubsection{Analysis of two  cases: matter in equilibrium and 
gravitational collapse}\label{analysis} 

We can now apply the above equations of motion to a variety of 
physical situations.  We choose two interesting cases, namely static 
matter 
in equilibrium and 
gravitational collapse of matter. 

\vskip 0.4cm 
{\noindent (a) Matter in equilibrium} 
\vskip 0.1cm 

To find a solution for a static shell we have to determine the radii 
where both the velocity of the shell and its acceleration are null, i. 
e.,  $\dot{R}=0$ and $\ddot{R}=0$, in the Eqs. 
(\ref{squaredspeedequation1}) 
(or, alternatively, (\ref{squaredspeedequation2})) and 
(\ref{accelerationequation}), respectively. Forcing $\dot{R}=0$ in 
(\ref{squaredspeedequation1}) 
(or, alternatively, (\ref{squaredspeedequation2})) yields 
\begin{equation} 
R_0^{d-3}\left[\frac12 m \,(\sqrt{f^2_+} +\sqrt{f^2_-})\,- 
\left( M_+ - M_- \right)\right] = 
\frac{\epsilon (Q_+^2 - Q_-^2)}{2 (d-3)}\,, 
\label{pressureraidal} 
\end{equation} 
where $R_0$ is the radius at which the 
shell is static, and $f^2_+$ and $f^2_-$ are evaluated 
at $R_0$. To be in equilibrium we have to impose, in addition, 
$\ddot{R}=0$, which when combined simultaneously with 
(\ref{pressureraidal}) gives the pressure 
necessary to hold the thin shell in equilibrium. This pressure is 
then given by 
\begin{eqnarray} 
P &=& - \frac{1}{(d-2)R_0^{d-3}\,\Omega_{d-2}\,\gamma_+\gamma_-}\left[ 
\frac{(Q^2_+ - Q^2_-)\,\gamma_+\gamma_-m}{2\,E\,R_0^{d-2}} 
-\frac{m^2}{4\,E}\,\left( \gamma_- 
\left.\frac{df^2_+}{dR}\right|_{R_0} + \gamma_+ 
\left.\frac{df^2_-}{dR}\right|_{R_0} \right)\right]\,, 
\label{pressuregeneral} 
\end{eqnarray} 
where the functions are evaluated at 
$R_0$, given by (\ref{pressureraidal}).  Thus, we obtain the pressure 
in terms of the parameters of the problem, such as $d$, $k$, $m$, 
$M_+$, $M_-$, $Q_+$, $Q_-$, and the 
radius of the static shell $R_0$, for a static configuration. 
As a particular simple case we may apply the above expressions to find 
the pressure necessary to hold the shell in static equilibrium in 
general relativity ($k=1,\,\chi=1$), without charge, and for zero 
cosmological constant and flat interior. 
This gives 
\begin{equation} 
P=\frac{(d-3)}{2(d-2)}\,\frac{G\,\,m^2} 
{\Omega_{d-2}\,R_0^{d-2}(R_0^{d-3}-m)}\,, 
\label{pressure1} 
\end{equation} 
where $R_0$ is the radius of the static shell, given by 
$R_0=\left(G\,m^2/[2(m-M)]\right)^{1/(d-3)}$, and $G$ is the Newton's 
constant for $k=1$, i.e., we have done $G_{k=1}\equiv G$, whatever 
the dimension $d$. In four dimensions, 
$d=4$,  Eq. (\ref{pressure1}) reduces to 
$ 
P=(G\,m^2)/(16\,\pi\,R_0^2(R_0-m))\,, 
\label{pressure2} 
$ 
with $R_0=G\,m^2/[2(m-M)]$, confirming the result given in \cite{Kuch}. 

\vskip 0.4cm 
{\noindent (b) Gravitational collapse } 
\vskip 0.1cm 

In this analysis of gravitational collapse in Lovelock gravity 
coupled to Maxwell electromagnetism, we 
first study a simple example of charged dust matter 
collapsing into an empty interior and 
then prove cosmic censorship in the generic case of shell collapse 
into an interior free of naked singularities. 
An analysis of gravitational expansion, not 
worked out here, follows straightforwardly by performing a time reversal 
operation on the collapsing solutions. 

\vskip 0.3cm 
{\noindent \it (i) Gravitational collapse of dust matter in 
an empty interior:} 
\vskip 0.1cm 

In order to have an idea of the main features of gravitational 
collapse in  Lovelock gravity 
coupled to Maxwell electromagnetism, 
rather than analyzing the full problem, which can be daunting since 
there are a great number of parameters to play with, we investigate 
the simplest problem, namely, of gravitational collapse of charged 
dust matter 
into an empty interior.  Even then, in this case the task of studying it 
in full detail is enormous, since one has first to set the dimension 
of the spacetime, the dimension of the $k$ parameter, and then all 
the other parameters. So we resort to study it somewhat generically, and 
then give some particular examples in the diverse theories that 
we chose to study. 
The aim is to show that gravitational collapse is possible in 
this subset of theories 
of Lovelock gravity, reinforcing some similarities 
which this has with general relativity, as well as showing some 
new features.  So, in the following we 
analyze the case where the matter is composed of dust particles with 
$P=0$, in which case $m$ is a constant and can be identified with the 
shell's rest mass, and where the interior is empty, in which case the 
function $f_-(R)$ is given through $f^2_-(R)=R^2/l^2+1$. 

Before proceeding, it is necessary to emphasize that the Lovelock 
theory we are analyzing is separable into two branches: (i) the 
$d-2k-1>0$, which can have even and odd dimensions, and includes 
general relativity when $k=1$ (any $d$), includes the Born-Infeld case 
of the dimensionally continued theory when $k=[\frac{d-1}{2}]$, and 
includes other generic cases, and (ii) the $d-2k-1=0$, which can only 
have odd dimensions and is precisely the Chern-Simons case of the 
dimensionally continued theory. We study each branch in turn. 

\vskip .4cm 
\noindent The branch $d-2k-1>0$ (general relativity when 
$k=1$ (any $d$), Born-Infeld when $k=[\frac{d-1}{2}]$, 
and other generic cases): 
\vskip .1cm 
\noindent The $d-2k-1>0$ branch allows theories in even and odd 
dimensions. 
For instance in $d=6$ one can have $k=1$ (general relativity) and 
$k=2$ (general relativity with a generalized Gauss-Bonnet term) 
theories, which in the case $k=2$ gives a Born-Infeld 
theory.   On the other hand, in 
$d=7$ one can also have theories with $k=1$ (general relativity) and 
$k=2$ (general relativity with a generalized Gauss-Bonnet term), 
but none of these is a Born-Infeld theory. 
Indeed, as already alluded to in Sec. \ref{LLSection}, the 
Born-Infeld case is the realization of this Lovelock theory in even 
dimensions, for which $k=[\frac{d-1}{2}]$. 

Considering thus an empty interior, 
with $M_-=0$ and $Q_-=0$, one has $f^2_-(R)=R^2/l^2+1$. 
Putting then $M_+=M$, $Q_+=Q$, and $\chi=(\pm 1)^{k+1}$ (see (\ref{chi})) 
one finds from Eqs. 
(\ref{squaredspeedequation1})-(\ref{squaredspeedequation2}) 
that 
the shell equation for $d-2k-1>0$ is, 
\begin{eqnarray} \dot{R}^2 &=& \left( \frac{M-\frac{\epsilon 
Q^2}{2\,(d-3)\,R^{d-3}}}{m}+\frac{m}{4\, \left(M-\frac{\epsilon 
Q^2}{2\,(d-3)\,R^{d-3}}\right)} \right.  \times \nonumber\\ &\times& 
\left. \chi \left(\frac{2G_k}{R^{d-2k-1}}\right)^{1/k} \left( 
M-\frac{\epsilon}{2\,(d-3)}\frac{Q^2}{R^{d-3}} \right)^{1/k} \right)^2 
- \nonumber \\ 
&-& \left(1+\frac{R^2}{l^2}\right)\,, 
\label{squaredspeedequationBI} 
\end{eqnarray} 
In order to have physical solutions, we have to 
demand $\dot{R}^2>0$. 

\begin{figure}[ht] 
\begin{center} 
\includegraphics[width=0.35\textwidth]{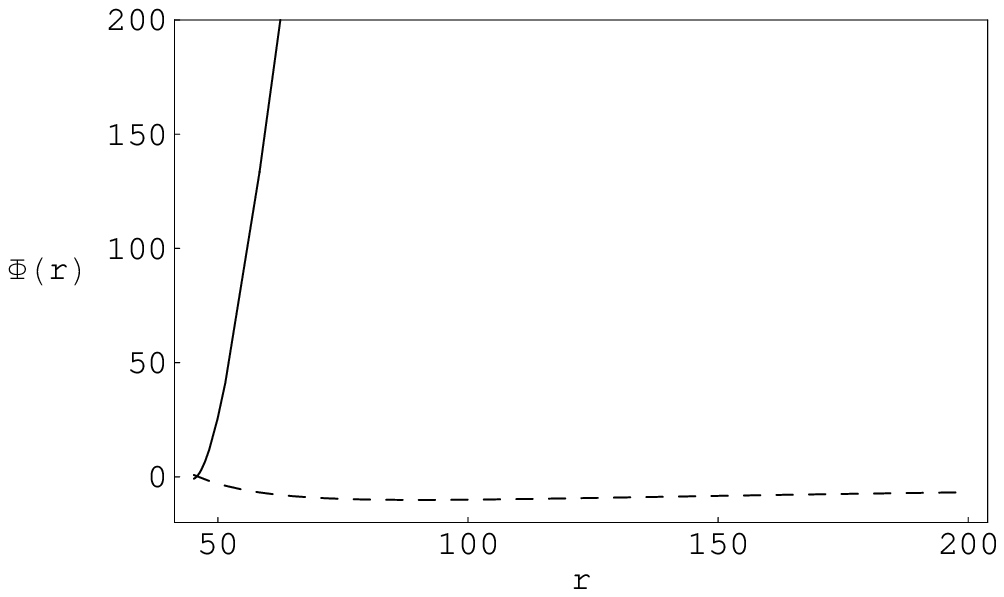} \quad 
\includegraphics[width=0.35\textwidth]{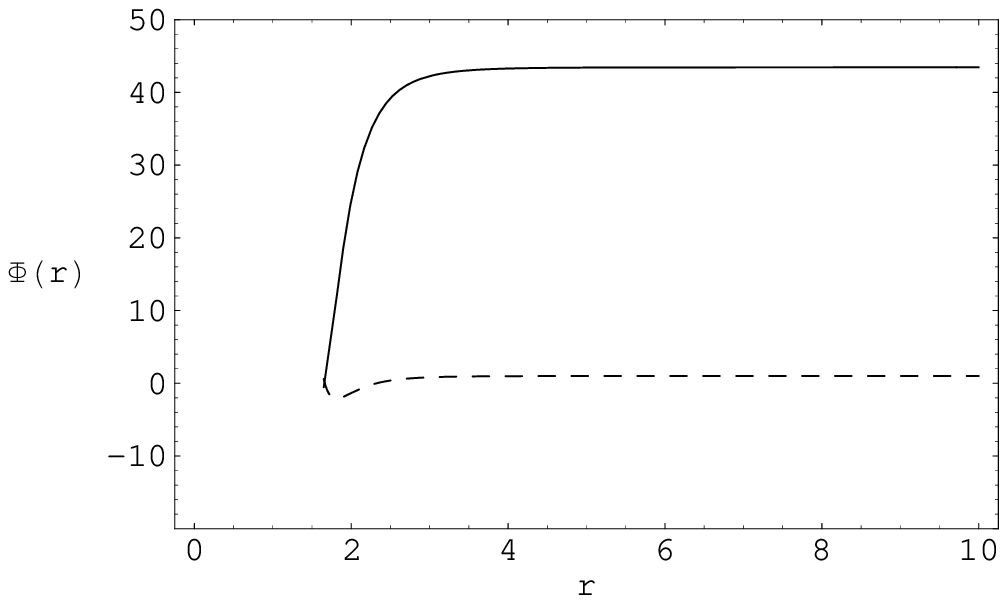} 
\end{center} \centerline{(a)} \centerline{} 

\begin{center} 
\includegraphics[width=0.35\textwidth]{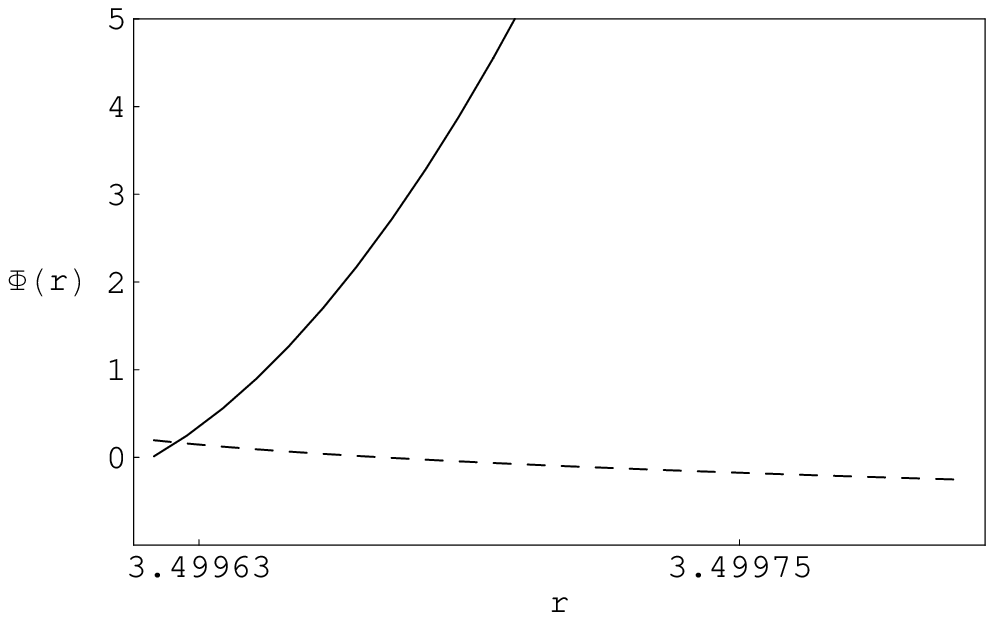} \quad 
\includegraphics[width=0.35\textwidth]{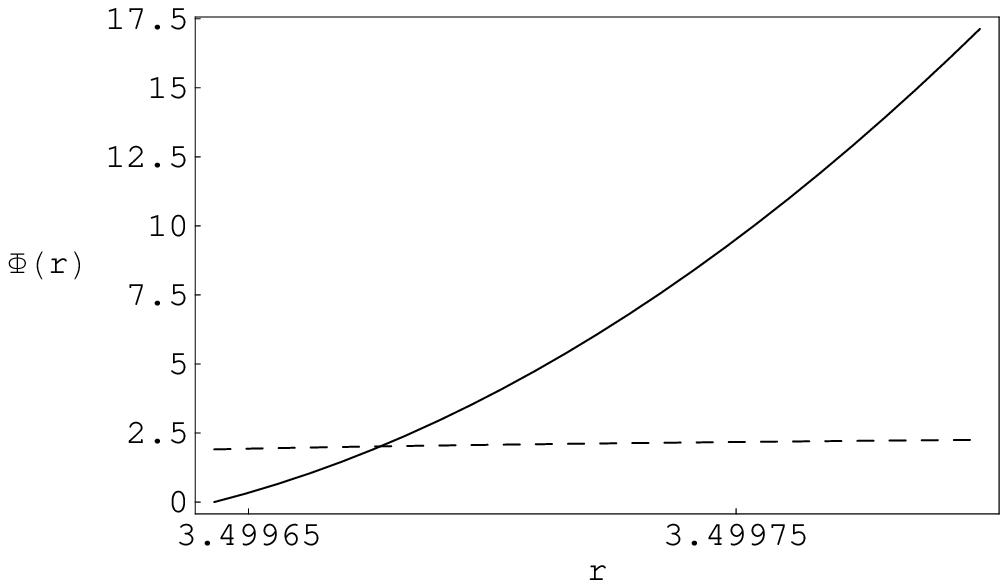} 
\end{center} \centerline{(b)} 

\begin{center} 
\includegraphics[width=0.35\textwidth]{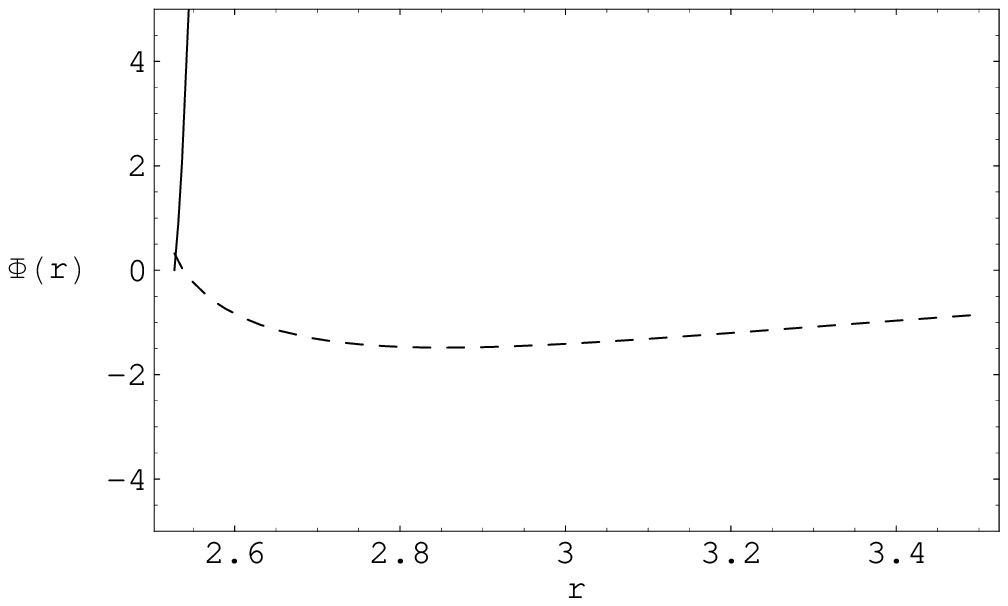} \quad 
\includegraphics[width=0.35\textwidth]{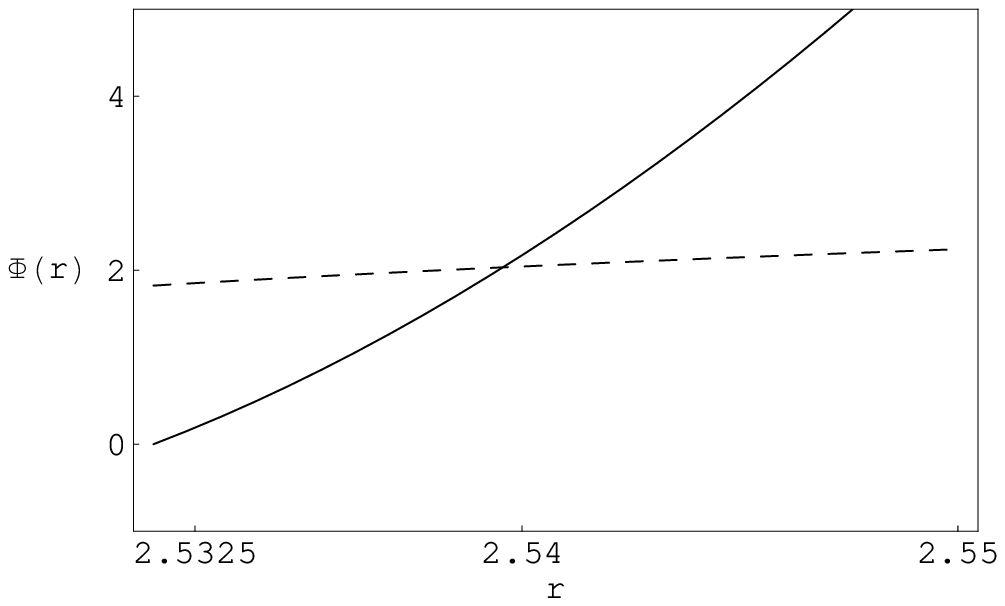} 
\end{center} \centerline{(c)} \centerline{} 

\caption{ 
In this figure the branch $d-2k-1>0$ (general relativity when $k=1$, 
Born-Infeld when $k=[\frac{d-1}{2}]$, and other generic 
cases) is studied in some particular instances.  In each plot 
the effective potential $\dot{R}^2$ as a function of $r$ (which gives 
the turning points and the allowed regions for the shell), 
and the metric function $f^2(r)$ as a function of 
$r$ (which gives the horizon formation, for several different values 
of the parameters) are displayed. 
The vertical axis 
$\Phi(r)$ represents both $\dot{R}^2(r)$ 
(full line) and $f_+^2(r)$ (dashed line). 
Note that the thin shell trajectory is allowed 
only in the region where $\dot{R}^2$ is positive.  In all the plots 
the respective Newton's constant is put equal to the unity, 
$G_k=1$, so that radii and energies  are measured in the 
respective Planck units ($\hbar=1$, $c=1$). 
(a) General relativity, $k=1$, left plot is for $d=4$ and right 
plot for $d=10$, with $\chi=1$ in both plots ($\chi=-1$ is not 
possible for $k$ odd), 
(b) Born-Infeld, $k=[\frac{d-1}{2}]$, left plot is for 
$\chi=1$ and right plot for $\chi=-1$, 
(c) General relativity with a generalized Gauss-Bonnet term, $k=2$, a 
particular case of $1<k<[\frac{d-1}{2}]$, left plot 
is for $\chi=1$ and right plot for $\chi=-1$.  See text for details.} 
\label{grfigure} 
\label{bifigure} 
\end{figure} 

The equation of the acceleration of the thin shell 
(\ref{accelerationequation}),  allows us to understand the forces 
acting on the shell in collapse. Expanding the $\gamma_\pm$  defined 
in (\ref{gammaplusminus}), with $\dot{R}^2$ replaced by the right 
hand side of 
(\ref{squaredspeedequation1}) or (\ref{squaredspeedequation2}), 
appropriately chosen, we have 
\begin{eqnarray} 
\gamma_\pm &=& \left| \frac{M-\frac{\epsilon 
Q^2}{2\,(d-3)\,R^{d-3}}}{m}\pm \frac{m}{4\,\left(M-\frac{\epsilon 
Q^2}{2\,(d-3)\,R^{d-3}}\right)} \right. \times \nonumber\\ &\times& 
\left. \chi \left(\frac{2G_k}{R^{d-2k-1}}\right)^{1/k} \left( 
M-\frac{\epsilon}{2\,(d-3)}\frac{Q^2}{R^{d-3}} \right)^{1/k} 
\right|\,. 
\label{gammasBI} 
\end{eqnarray} 
Then Eq. (\ref{accelerationequation}) for an empty 
interior turns into 
\begin{eqnarray} 
\ddot{R} 
&=& -\frac{2\,R}{l^2}+\frac{Q^2\gamma_+ \gamma_- 
}{2\,E\, \,R^{d-2}}-\, \chi\,\frac{m}{2\,E} \times \left( \gamma_- \, 
\frac1k\left(\frac{2\,G_k\,M}{R^{d-2k-1}}-\frac{\epsilon\,G_k}{d-3} 
\frac{Q^2}{R^{2(d-k-2)}}\right)^{\frac{1-k}{k}}\right. \nonumber \\ && 
\times \left. \left( 
\frac{2(d-2k-1)\,G_k\,M}{R^{d-2k}}-\frac{2\,(d-k-2)\epsilon\,G_k}{d-3} 
\frac{Q^2}{R^{2(d-k)-3}} \right)\right)\,, 
\label{acceleration2} 
\end{eqnarray} 
where the functions $\gamma_\pm$ are given in 
(\ref{gammasBI}), and we have used Eq. (\ref{shellequation2}). 
The first term on the right hand side of Eq. 
(\ref{acceleration2}) is proportional to the radius of the thin shell, 
meaning that this term dominates the acceleration 
for large values of $R$, it tends to $-\infty$ as 
$R\rightarrow\infty$ dominating all the others, 
and 
since the term is negative 
it points towards the center. The term proportional to the 
square of  the electric charge $Q^2$ includes a 
$\dot{R}^2$ terms and a $|\dot{R}|$ term (due to the 
product $\gamma_-\gamma_+$). This means that 
the charge term is positive for a thin 
shell, works to push the shell to higher radii, and it tends to zero 
as $R\rightarrow\infty$.  The third term has several terms 
included, like gravitational and viscosity forces per unit mass 
terms. 
A simple example where the terms in 
Eq. (\ref{acceleration2}) are drastically reduced 
is when we put 
$d=4,\,k=1,\,l^2\rightarrow\infty,\,Q=0$, i.e., the case of an 
uncharged shell collapse in usual general relativity 
in four dimensions. Then we get for the acceleration of the shell, 
$ 
\ddot{R}= - M/R^2+ m^2/(4\,R^3)\,. 
$ 
The term $m^2/(4\,R^3)$ is a correction to the 
Newtonian dynamics, representing a self gravitational potential energy. 

Since, in general, the character $\chi$ has two values, 
$\chi=\pm1$, for the branch we are analyzing, we 
study separately both cases. Note first that for $\chi=1$ one has from 
(\ref{chargedsolution4}) that the shell spacetime solution has a black 
hole character (i.e., for certain choices of the parameters 
one has a black hole solution), whereas for $\chi=-1$ one has from 
(\ref{chargedsolution4}) that the shell spacetime solution has a naked 
singularity character (i.e., there are no possible choice of 
parameters that give a black hole). 
So, since when $k$ is odd (such as in general 
relativity where $k=1$) the character 
$\chi=1$ for sure, odd $k$ allows black hole 
character solutions only, and since when $k$ is even (such as in general 
relativity with a Gauss-Bonnet term where $k=2$) the 
character $\chi$ can have both 
values $\pm1$, even $k$ can have both black hole and naked singularity 
character solutions. 

For $\chi=1$ the spacetime has a black hole character. 
When the shell collapses it will pass through its 
horizon radius, defined by the equation $f_+^2(r_{h})=0$. So 
for $\chi=1$  shell collapse 
implies the formation of a black hole.  To see this more clearly, 
note that from Eq. (\ref{chargedsolution4}) the 
relation $f_+^2(r_{h})=0$ defining the horizon turns into 
$\left( 1+r_{h}^2/l^2 \right) = 
\left(2G_k/r_{h}^{d-2k-1}\right)^{1/k} \left( 
M-(\epsilon\,Q^2)/(2\,(d-3)r_{h}^{d-3}) \right)^{1/k}.$ 
One can show by inspection that the horizon 
radius is always larger than the turning radius 
$r_t$, $r_{h}> 
r_t$. So in this type of Lovelock gravity the bounce is always 
inside the event horizon, which implies the shell expands 
into another universe, a result already obtained in 
pure general relativity \cite{GL}. 

For $\chi=-1$ the spacetime has a naked singularity character.  We can 
now give an argument which shows that cosmic censorship holds in the 
case of $d-2k-1>0$, with an empty interior.  The spacetime would be 
singular if the shell would hit $r_{e}$, given in (\ref{electricradius}). 
Through the shell equation, Eq. (\ref{shellequation2}), we see that 
$M-(\epsilon\,Q^2)/(2\,(d-3)R^{d-3}) > 0\,. 
\label{electricradiusBI} $ 
However, the singularity at $r_{e}$ 
makes the right hand side of the shell Eq. (\ref{shellequation2}) 
zero. This shows that the radius of this singularity is out of the 
region of validity of the shell equation, preventing the collapse of 
the charged shell to form a naked singularity, and validating here the 
cosmic 
censorship hypothesis.  Note that this is only applicable to $k$ even. 

In Figure \ref{grfigure} the branch $d-2k-1>0$ (general relativity 
when $k=1$, Born-Infeld when $k=[\frac{d-1}{2}]$, and other generic 
cases) is studied in some instances.  In each plot the 
effective potential $\dot{R}^2$ as a function of $R$ (which gives the 
turning points and the allowed region for the shell motion), and the 
metric 
function $f^2(R)$ as a function of $R$ (which gives the horizon 
formation), for several different values of the parameters, are shown. 
The vertical axis 
$\Phi(r)$ in the figure represents both $\dot{R}^2(r)$ 
(full line) and $f_+^2(r)$ (dashed line). 
The thin shell trajectory is allowed only in the region where 
$\dot{R}^2$ is positive.  In all the plots we have made the respective 
Newton's constant equal to the unity, $G_k=1$, so that radii and energies 
are measured in the respective Planck units ($\hbar=1$, 
$c=1$). We have not plotted $\ddot R$ since it yields a totally different 
scale in the vertical axis, rendering the other two functions, 
$\dot{R}^2$ and $f^2(R)$, almost invisible. 
We now look at each choice of plot in turn. 
(a) General relativity, $k=1$. Left plot: 
$d=4$, $l=\infty$, $M=1000$, 
$Q=300$, $m=20$. Right plot: $d=10$, $l=\infty$, $M=200$, $Q=300$, $m=30$ 
(since $k$ is odd, one has $\chi=1$ for both plots). More 
specifically, the left plot is the 
usual general relativity with $d=4$ (which incidentally is also a 
Born-Infeld type theory), and the right plot is general 
relativity with $d=10$.  There is always collapse to a black hole, a 
result confirmed by both plots. Indeed, since $\dot{R}^2$ is nonzero 
at the horizon, where $f_+^2(r)=0$, the shell passes smoothly through 
the horizon itself.  So there is no bounce outside the horizon, the 
bounce being inside the horizon, and into another universe. For 
example, in $d=4$ the value of the 
radius of the bounce, or turning point, inside the horizon is given by 
$r_t=(m^2-Q^2)/(2(m-M))=45.7$ in the units 
defined above. This agrees with 
the expressions given in \cite{GL}, where using Israel's 
formalism \cite{Isr}, as opposed to the Hamiltonian methods 
used here, a full study of the 
equation of motion and of the 
turning points in $d$ dimensional general relativity was performed. 
(b) Born-Infeld, $k=[\frac{d-1}{2}]$. 
Left plot: $d=10$, $k=4$, $\chi=1$, $l=\infty$, $M=10000$, $Q=30000$, 
$m=1$. Right plot: $d=10$, $k=4$, $\chi=-1$, $l=\infty$, $M=10000$, 
$Q=30000$, $m=1$. More specifically, the left plot is for 
$\chi=1$, and we find there is always collapse to a black hole, 
since $\dot{R}^2$ is nonzero at the horizon, where $f_+^2(r)=0$, and 
the shell passes smoothly through the horizon itself.  There is a bounce 
inside the horizon into another universe. 
The right plot is for $\chi=-1$, with the 
same choices of the other parameters as for $\chi=1$. In this 
case the external spacetime has a naked singularity character. 
Here there is no horizon because the metric function $f_+^2(r)$ has no 
zero.  There is a zero of $\dot{R}^2$, which limits the region 
$\dot{R}^2>0$, the region where the shell has its trajectory. The 
zero is larger than the 
radius at which there is a singularity.  Thus the collapse never forms 
a naked singularity. 
(c) General relativity with a generalized Gauss-Bonnet term, $k=2$, a 
particular case of $1<k<[\frac{d-1}{2}]$, 
(here the intermediate cases, ($1<k<[\frac{d-1}{2}]$), where $k$ is 
neither 
minimum (general relativity), nor maximum (Born-Infeld), are  studied). 
Left plot: 
$d=10$, $k=2$, $\chi=1$, $l=\infty$, $M=1000$, 
$Q=300$, $m=20$. Right plot: $d=10$, $k=2$, $\chi=-1$, $l=\infty$, 
$M=1000$, 
$Q=300$, $m=20$. The choice of the coefficients $\alpha_p$ are given 
in Eq. (\ref{alphaconstants}). 
More 
specifically, the left plot is for 
$\chi=1$, and we see 
there is always collapse to a black hole, since $\dot{R}^2$ is 
nonzero at the horizon, where $f_+^2(r)=0$, and the shell crosses the 
horizon smoothly. There is also a bounce inside the horizon. 
The right plot 
is for $\chi=-1$, where 
the external spacetime has a naked singularity character. There is 
no horizon, because there are no zeros of $f_+^2(r)$. The region where 
$\dot{R}^2>0$, where the shell has its trajectory, is limited on the 
left by the zero of $\dot{R}^2$. The zero of $\dot{R}^2$ is larger 
than the radius at which there 
is a singularity. There is thus no collapse to form a naked 
singularity. In all the figures (a), (b) and (c) we have shown 
bouncing solutions only, although one could easily find parameters 
for which the collapsing shell suffers no bounce and goes 
all the way down to the singularity. 
Note that a general set up for the study of the turning points 
and the shell's trajectory in a Carter-Penrose diagram for 
the causal structure 
(with a corresponding careful analysis of the normal to the 
shell along the trajectory), 
as was done analytically for $d$ dimensional general relativity in 
\cite{Boul,GL}, 
could also be performed here case to case, i.e., giving $d$ and 
$k$. However, this is beyond the scope of this work. 

\vskip .4cm 
\noindent The branch $d-2k-1=0$ (Chern-Simons): 
\vskip .1cm 
\noindent 
The $d-2k-1=0$ branch implies $d$ is odd always. Moreover the theory 
is of Chern-Simons type. The $d-2k-1>0$ branch and the Chern-Simons 
$d-2k-1=0$ branch, are quite distinct. First, in this latter branch, 
in this charged setting, there is no general relativity ($k=1$), because 
the 
theory is defined only for $d>3$, and thus $k\geq2$ 
(however in the uncharged case the theory is well defined 
for $d=3$, yield three dimensional general relativity, 
see \cite{BTZ1}).  Second, for the $d-2k-1>0$ branch one has that 
the vacuum mass is $M_-=0$, whereas for the $d-2k-1=0$ Chern-Simons 
branch the vacuum mass is $M_-=-(2\,G_k)^{-1}$. From 
(\ref{chargedsolution4}) 
and (\ref{gfunction}) one finds that the interior for the Chern-Simons 
theory with $M_-=-(2\,G_k)^{-1}$ is characterized by 
$f^2_-(R)=R^2/l^2+1$ also. 
\begin{figure}[h] \begin{center} 
\includegraphics[width=0.35\textwidth]{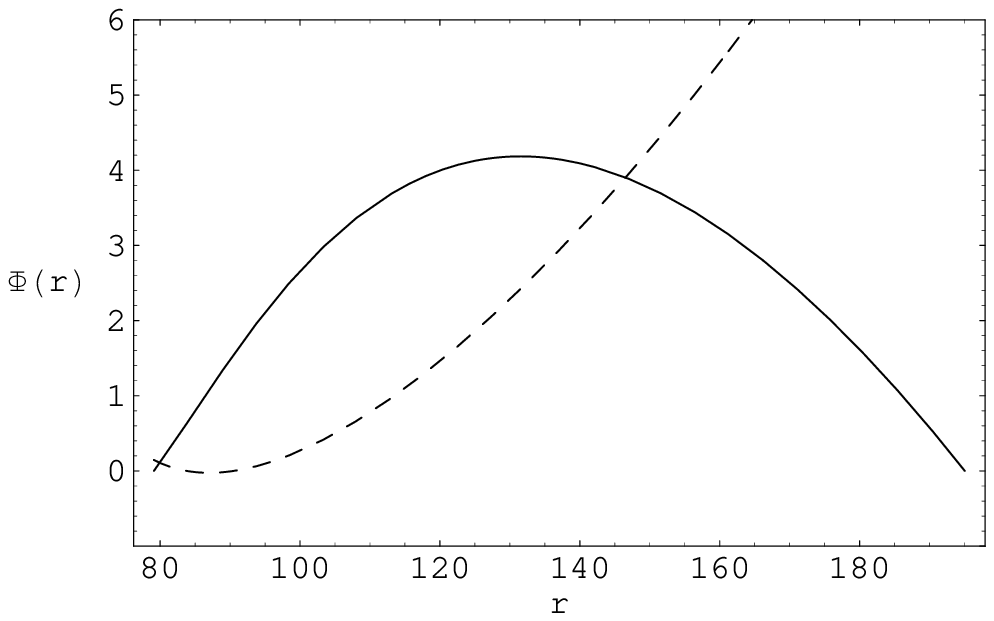} \quad 
\includegraphics[width=0.35\textwidth]{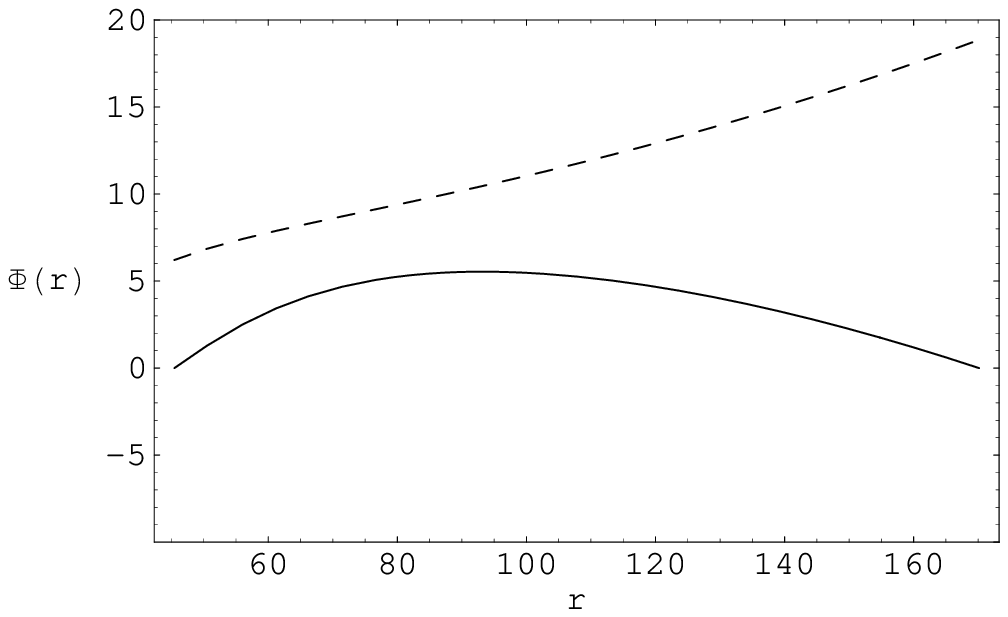} 
\end{center} 
\caption{ 
In this figure the branch $d-2k-1=0$ (Chern-Simons, $d$ always 
odd) is studied in some particular instances. 
In each plot 
the effective potential $\dot{R}^2$ as a function of $r$ (which gives 
the turning points and the allowed regions for the shell), 
and the metric function $f^2(r)$ as a function of 
$r$ (which gives the horizon formation, for several different values 
of the parameters) is displayed. 
The vertical axis 
$\Phi(r)$ represents both $\dot{R}^2(r)$ 
(full line) and $f_+^2(r)$ (dashed line). 
Note that the thin shell trajectory is allowed 
only in the region where $\dot{R}^2$ is positive.  In the plots 
the respective Newton's constant is put equal to the unity, 
$G_k=1$, so that radii and energies  are measured in the 
respective Planck units ($\hbar=1$, $c=1$). 
Left plot is for $k=2$, $d=5$ and $\chi=1$, and right plot 
is for $k=2$, $d=5$ and $\chi=-1$. See text for details} 
\label{csfigure} 
\end{figure} 

    From Eq. (\ref{shellequation2}) the shell equation for 
$d-2k-1=0$, $M_-=-(2\,G_k)^{-1}$, $M_+=M$, $Q_-=0$, $Q_+=Q$, and 
$\chi=(\pm 1)^{k+1}$ is, 
\begin{eqnarray} \dot{R}^2 &=& \left( \frac{M+(2\,G_k)^{-1}- 
\frac{\epsilon Q^2}{2\,(d-3)\,R^{d-3}}}{m}+\frac{m}{4\, 
\left(M+(2\,G_k)^{-1}-\frac{\epsilon Q^2}{2\,(d-3)\,R^{d-3}}\right)} 
\right. \times \nonumber\\ &\times& \left. \chi 
\left(2G_k\right)^{1/k} \left( 
M+(2\,G_k)^{-1}-\frac{\epsilon}{2\,(d-3)}\frac{Q^2}{R^{d-3}} 
\right)^{1/k} \right)^2 - \nonumber \\ &-& 
\left(1+\frac{R^2}{l^2}\right)\,. 
\label{squaredspeedequationCS} 
\end{eqnarray} 
In order to have physical solutions, we have to 
demand $\dot{R}^2>0$. 

Since, in general, $\chi=\pm1$, we 
study separately both cases. Again note first that for $\chi=1$ one has 
from 
(\ref{chargedsolution4}) that the shell spacetime solution has a black 
hole character, whereas for $\chi=-1$ one has from 
(\ref{chargedsolution4}) that the shell spacetime solution has a naked 
singularity character.  So, since when 
$k$ is even (such as in general 
relativity with a generalized 
Gauss-Bonnet term, where $k=2$ and $d=5$) $\chi$ can have both 
values $\pm1$, even $k$ can have both black hole and naked singularity 
character solutions, and since when 
$k$ is odd (such as $k=3$, and 
$d=7$) $\chi=1$ for sure, odd $k$ allows solutions of black hole 
character only. 

For $\chi=1$ (so $k$ can be odd or even) 
the horizon is located at the radius 
$r_{h}$ given implicitly by the equation (see 
(\ref{chargedsolution4})) 
$\left(1+r_{h}^2/l^2\right) = 
\left(2G_k\right)^{1/k}\left( M+(2G_k)^{-1}- 
(\epsilon\,Q^2)/(2\,(d-3)r_{h}^{d-3}) \right)^{1/k}.$ 
When the parameters $m$, 
$M$, $Q$, $l$ yield a solution for the collapsing shell 
then the shell will pass through its own 
event horizon at $r_{h}$ and form a black hole. 

For $\chi=-1$ (so $k$ can be even only), the shell solution has a 
naked singularity character.  We can again give an argument to show 
that cosmic censorship in the particular case of Chern-Simons ($d-2k-1=0$) 
with an empty interior holds.  Indeed, from Eq. (\ref{shellequation2}) one 
finds that the following relation holds 
$M+(2\,G_k)^{-1}-(\epsilon\, 
Q^2)/(2\,(d-3)\,R^{d-3})>0\,.$ 
Noting that $r_{e}$, given in 
(\ref{electricradius}), makes the right hand side zero, this means that 
this singularity is unattainable through gravitational collapse. Thus 
cosmic censorship holds in charged collapse in a Chern-Simons type 
theory. This is in contrast to uncharged collapse in the same 
theory where naked singularities can form \cite{CCS}. 

Finally we write the equation of the acceleration of the thin shell, 
which allows us to understand the forces acting on the collapsing 
shell. Expanding the $\gamma_\pm$  defined 
in (\ref{gammaplusminus}), with $\dot{R}^2$ replaced by the right 
hand side of 
(\ref{squaredspeedequation1}) or (\ref{squaredspeedequation2}), 
appropriately chosen, we have 
\begin{eqnarray} 
\gamma_\pm &=& \left| \frac{M+(2\,G_k)^{-1}- 
\frac{\epsilon Q^2}{2\,(d-3)\,R^{d-3}}}{m}\pm\frac{m}{4\, 
\left(M+(2\,G_k)^{-1}-\frac{\epsilon Q^2}{2\,(d-3)\,R^{d-3}}\right)} 
\right.  \times \nonumber\\ &\times& \left. \chi 
\left(2G_k\right)^{1/k} \left( 
M+(2\,G_k)^{-1}-\frac{\epsilon}{2\,(d-3)}\frac{Q^2}{R^{d-3}} 
\right)^{1/k} \right|\,. 
\label{gammasCS} 
\end{eqnarray} 
As before, the constraint radius marks the 
lower bound of validity of the shell's equations. Then the Eq. 
(\ref{accelerationequation}) for an empty interior turns into 
\begin{eqnarray} \ddot{R}&=& \frac{Q^2\gamma_+ \gamma_- }{2\,E\, 
\,R^{d-2}} - \frac{m}{2\,E} \times \left( \gamma_- 
\left[\frac{2\,R}{l^2}-\chi\, 
\frac{2}{d-1}\left(2\,G_k\,M+1-\frac{\epsilon\,G_k}{d-3} 
\frac{Q^2}{R^{d-3}}\right)^{-\frac{d-3}{d-1}}\right.\right. \nonumber 
\\ && \times \left.\left. \left(\epsilon\,G_k\, \frac{Q^2}{R^{d-2}} 
\right)\right] + \gamma_+\,\frac{2\,R}{l^2} \right) \nonumber \\ &=& 
-\frac{2\,R}{l^2}+\frac{Q^2\gamma_+ \gamma_- }{2\,E\, \,R^{d-2}}+\, 
\chi\,\frac{m}{2\,E} \times \left( \gamma_- \, 
\frac{2}{d-1}\left(2\,G_k\,M+1-\frac{\epsilon\,G_k}{d-3} 
\frac{Q^2}{R^{d-3}}\right)^{-\frac{d-3}{d-1}}\right. \nonumber \\ && 
\times \left. \left(\epsilon\,G_k\, \frac{Q^2}{R^{d-2}} 
\right)\right)\,, 
\label{acceleration3} 
\end{eqnarray} 
where the functions $\gamma_\pm$ are given in (\ref{gammasCS}), and we 
used Eq. (\ref{shellequation2}) in the second equality. 
For uncharged collapse $Q=0$ the acceleration is always 
negative, as can be easily checked from Eq. (\ref{acceleration3}). 
In the  charged case $Q\neq0$, there is no 
definite sign on the right hand side of Eq. 
(\ref{acceleration3}), it depends on the radius at 
which the shell is located. Thus, re-expansion, after a collapsing 
phase, is allowed. 

                   In Figure \ref{csfigure} we show the effective 
potential 
$\dot{R}^2$ as a function of $R$ (which gives the turning points and 
the allowed regions for the shell), and the metric function $f^2(R)$ 
as a function of $R$ (which gives the horizon formation), for several 
different values of the parameters. 
The vertical axis 
$\Phi(r)$ in the figure represents both $\dot{R}^2(r)$ 
(full line) and $f_+^2(r)$ (dashed line). 
The thin shell 
trajectory is allowed only in the region where $\dot{R}^2$ is 
positive.  In all the plots we have made the respective Newton's 
constant equal to the unity, $G_k=1$, so that radii, energies and 
forces are measured in the respective Planck units ($\hbar=1$, 
$c=1$). Left plot: $d=5$, 
$k=2$, $\chi=1$, $l=50$, $M=20$, $Q=610$, $m=5$. 
Right plot: 
$d=5$, $k=2$, $\chi=-1$, $l=50$, $M=20$, $Q=300$, $m=5$. 
More specifically, 
the left plot shows a region limited by two zeros of 
$\dot{R}^2$, the region where a trajectory of a thin shell is 
possible. The horizon, given by the larger zero of $f^2(R)$, is inside 
the region where $\dot{R}^2>0$, which means that the shell passes 
smoothly through the horizon. The shell then suffers a bounce inside 
the horizon into another universe. The plot on the right shows a 
naked singularity character exterior spacetime, which means that there is 
no zero of $f^2(R)$, hence no horizon. There is, however, no collapse 
to a naked singularity, because the region where $\dot{R}^2>0$ is 
limited by two zeros, the smaller being larger than the constraint 
radius. Thus the shell bounces between two extreme values of $R$, and 
so does not collapse at all.  In the figure we have shown bouncing 
solutions only, although as before one could produce totally 
collapsing solutions.  As in the $d-2k-1>0$ branch, a general set up 
for the study of the turning points and the shell's trajectory in a 
Carter-Penrose diagram for the causal structure 
could be performed here, but this will not be done. 

\vskip 0.3cm 
{\noindent \it (ii) Cosmic censorship:} 
\vskip 0.1cm 

Given the experience we have acquired with the above examples, 
one can now study cosmic censorship directly from 
Eq. (\ref{shellequation2}) or (\ref{equation3}). 
Cosmic censorship holds if no naked singularity 
forms from gravitational collapse in an initially 
nonsingular spacetime or if no naked singularity 
forms from gravitational collapse 
in a spacetime initially containing 
a black hole. We assume $Q_+\neq0$ and $Q_-\neq0$. 
There are three such cases in our study: 
first, both spacetime regions on each side of the collapsing shell 
have character $\chi=1$, and the interior solution is a black hole 
spacetime; 
second, both spacetime regions on each side of the collapsing shell 
have character $\chi=1$, and the interior solution is an empty vacuum 
spacetime; 
and third, both spacetime regions on each side of the collapsing shell 
have character $\chi=-1$, and the interior solution is an empty vacuum 
spacetime. 
On all these cases cosmic censorship holds. 
For the first case, that both spacetime regions on each side of the 
collapsing shell have character $\chi=1$, and the interior solution is 
a black hole spacetime, given the above dynamic equation one can 
directly prove that the collapse of a charged shell in such a 
background never yields a naked singularity spacetime.  To start with 
we have a shell collapsing into an interior black hole spacetime. 
This means that the $\chi$ term on the left of (\ref{equation3}) is 
positive (otherwise the interior would not be that of a black hole 
spacetime).  From Eq.  (\ref{shellequation2}) it is known that 
$g_{k+}(R)^k-g_{k-}(R)^k>0$, because this expression is the right hand 
side of (\ref{shellequation2}). If then the shell is collapsing onto 
an existing black hole, and it is such that the result would be a 
naked singularity, i.e., the collapsing shell is sufficiently 
overcharged, then $f_-^2(r_{h})=0$ for a certain $r_{h}$, and 
$f_+^2(r)$ would always be larger than zero (because in the exterior 
spacetime there would be no horizon).  It is then clear that the term 
$(\gamma_--\gamma_+)$ is negative if the shell reaches 
$r_{h}$. However, as we are working in the asymptotically outside 
region, then $g_k(r)>0$. This means that the signal of $g_k(r)$ is the 
same as that of $g_k(r)^k$, and so the signal of 
$g_{k+}(R)^k-g_{k-}(R)^k$ is the same as that of 
$g_{k+}(R)-g_{k-}(R)$. This results in the fact that the left hand 
side of (\ref{equation3}) is positive at the horizon of the inner 
black hole, defined by $f_-^2(r_{h})=0$, and that the right hand side 
of the same Eq. (\ref{equation3}) is negative, which implies a 
contradiction. 
For the second case, that both spacetime regions on each side of the 
collapsing shell have character $\chi=1$, and the interior solution is 
an empty vacuum spacetime, one can treat it as a limiting case of the 
first case, as the mass of the interior black hole goes to zero, or 
directly using similar arguments as above, with the result that a 
collapsing shell, if undercharged forms a black hole, if overcharged 
has a bounce back from where it came. 
For the third case, that both spacetime regions on each side of the 
collapsing shell have character $\chi=-1$, and the interior solution 
is an empty vacuum spacetime, one can write Eq. (\ref{equation3}) as 
$m= - g_{k+}(R)^{k-1}\,R^{d-2k-1}\,(\gamma_- -\gamma_+)\,. 
\label{equation3cosmic3}$ One sees that when the shell approaches 
$r_{e}$, its right hand side approaches zero, whereas the left hand 
side approaches a finite value.  So there is a bounce and the shell 
never collapses to a singularity. 
These three cases prove the result, and yield cosmic censorship in 
charged backgrounds in the Lovelock theory we are studying 
coupled to Maxwell electromagnetism. 
We now briefly comment on the uncharged case $Q_+=Q_-=0$.  One can 
prove that in this particular case, in the $d-2k-1>0$ branch one has 
no formation of naked singularities, whereas in the $d-2k-1=0$ 
uncharged branch there is formation of naked singularities, 
so that one can say 
that electrical charge acts really as a cosmic censor. 

Thus, a charged shell never develops a naked singularity. This is 
rather like having a shell with some angular momentum, a situation 
which is much harder to study directly in theories with $d>3$. 
Nonetheless, the special case of an uncharged collapsing shell in 
$d=3$ with angular momentum was studied in \cite{CriOl,Ol}, with the 
result that indeed the angular momentum also prevents the formation of 
a naked singularity. 

\section{Conclusions and physical implications} 
\label{section4} 

There are two main conclusions from this work.  One conclusion is that 
the Hamiltonian formalism is a powerful method to treat in a unified 
way spacetimes composed of several pieces, such as several vacua and 
thin shells, in theories much more complicated than general relativity 
such as the subset of theories derived from Lovelock gravity coupled 
to Maxwell electromagnetism we have studied.  The other conclusion is 
that, when the spacetimes in question have the same character of those 
spacetimes provided by general relativity $(\chi=1)$, the collapse of 
the thin shells in the backgrounds, black hole or otherwise, of each 
different type of Lovelock theory is in many ways similar to the 
collapse in general relativity itself, and when the spacetimes in 
question have the opposite character $(\chi=-1)$, some other new 
features appear.  This in turn has the following 
physical implications: if indeed there 
are extra dimensions with a 
relative large size, as exposed in the Introduction, then the new and the 
old features, when confronted with experimental data, can provide the 
signature to the uncovering not only of the actual spacetime dimension 
$d$, but also of the value of the parameter $k$, i.e., of which 
particular Lovelock gravity nature picks up at the appropriate scales, 
whether be it general relativity ($k=1$, any $d$), Born-Infeld 
($k={\rm maximum}$, even $d$), Chern-Simons ($k={\rm maximum}$, odd 
$d$), or other generic gravity (other $k$, any $d$). Of course, a full 
quantum treatment, or even a semiclassical approximation, would be 
much more appropriate for this kind of questions, but for Lovelock 
type gravities the technical difficulties are exponentiated easily, 
and thus it is advisable to start up with a classical analysis, as we 
did here. 

\section*{Acknowledgments} 

This work was partially funded by Funda\c c\~ao para a Ci\^encia e a 
Tecnologia (FCT) of the Ministry of Science, Portugal, through project 
POCTI/FIS/57552/2004.  GASD is supported by grant SFRH/BD/2003 from 
FCT. SG was supported by NSFC Grants No. 10605006 and No. 10373003 and 
the Scientific Research Foundation for the Returned Overseas Chinese 
Scholars, State Education Ministry. JPSL thanks Observat\'orio Nacional 
do Rio de Janeiro for hospitality.

\end{document}